\documentclass[aps,floatfix,eqsecnum,showpacs,preprint,tightenlines]{revtex4}
\usepackage{epsfig}
\usepackage{amsmath}

\def\e{\kern+.6ex\lower.42ex\hbox{$\scriptstyle \iota$}\kern-1.20ex e}

\newcommand{\pipi}{$\pi-\pi\;$}
\newcommand{\pirho}{$\pi-\rho\;$}
\newcommand{\rhorho}{$\rho-\rho\;$}

\begin{document}

\title{
The Tucson-Melbourne Three-Nucleon Force in the automatized 
Partial Wave Decomposition
}

\author{R.~Skibi\'nski$^1$}
 
\author{J.~Golak$^1$}

\author{K.~Topolnicki$^1$}

\author{H.~Wita{\l}a$^1$}

\author{H.~Kamada$^2$}

\author{W.~Gl\"ockle$^3$}

\author{A.~Nogga$^4$}

\affiliation{$^1$M. Smoluchowski Institute of Physics, Jagiellonian
University, PL-30059 Krak\'ow, Poland}

\affiliation{$^2$ Department of Physics, Faculty of Engineering,
Kyushu Institute of Technology, Kitakyushu 804-8550, Japan}

\affiliation{$^3$Institut f\"ur Theoretische Physik II,
Ruhr-Universit\"at Bochum, D-44780 Bochum, Germany}

\affiliation{$^4$Forschungszentrum J\"ulich,
          Institut f\"ur Kernphysik, 
          Institute for Advanced Simulation
          and  J\"ulich Center for Hadron Physics, D-52425 J\"ulich, Germany}

\date{\today}

\begin{abstract}
A recently developed procedure for a partial wave decomposition of
a three-nucleon force is applied to the \pipi, \pirho and \rhorho
components  of the Tucson-Melbourne
three-nucleon potential. 
 The resulting matrix elements
for the \pipi and
\pirho components
are compared with the values obtained using 
the standard approach to the partial wave decomposition, in which 
 the \pirho expressions for the matrix elements  are also derived 
and presented. 
Several numerical tests and results for the triton binding energy and 
the correlation function prove the 
reliability and efficiency of the new method.
\end{abstract}
\maketitle \setcounter{page}{1}

\section{Introduction}

The Tucson-Melbourne (TM) three nucleon force 
(3NF)~\cite{Ellis.1985,Coon.1979,CoonGlockle.1981,Coon.1993}
is an important model of the
three-nucleon (3N) interaction. 
It consists of three parts
stemming from exchanges of
\pipi, \pirho and \rhorho mesons.
The main ingredient of the TM force,
the meson-nucleon scattering amplitude with the off-shell mesons, was
derived using the current algebra techniques.
This was done in~\cite{Coon.1979} 
and improved in~\cite{CoonGlockle.1981} for the \pipi part.
The \pirho and \rhorho contributions were derived 
in~\cite{Ellis.1985,Coon.1993}.
In~\cite{Friar.1999} the structure of the \pipi part of the TM 3NF was 
revisited to achieve a
 consistency  with the chiral symmetry and 
the modified force is known as the TM' model.

The effects of all terms on the triton binding energy were studied in
~\cite{Stadler.1995}. It turned out that the \pirho force acts 
repulsively for the $^3$H contrarily 
to the
\pipi interaction and combining them leads to the $^3$H binding
energy close to the experimental value. The $\rho-\rho$ force
has only a small influence on the triton binding energy.
A similar behaviour was observed for scattering observables in 
the three-nucleon system~\cite{Witala.1995}:
the largest effects came from the dominant \pipi part and the
influence of the \pirho part was smaller and in the opposite direction.
The \rhorho contribution proved to be much
smaller and practically negligible.
However, the results of Refs.~\cite{Stadler.1995, Witala.1995} were
based  on  partial waves
restricted to the total angular momenta in the two-nucleon subsystem $j \leq 2$.
Thus conclusions of~\cite{Witala.1995}
are valid only in a low energy domain of the three-nucleon continuum.
For higher energies, where more partial waves are required to achieve
convergence,  
only the \pipi part of the TM force was used,
 (see e.g.~\cite{Witala.2001,Zolnierczuk}). While 
the inclusion of this main component of the TM 3NF improves
the description of many scattering
observables, some serious
discrepancies with data remain and
they become larger at higher energies.
One of the possible explanations for this disagreement is a lack of
shorter-range parts of the 3NF in those  calculations, what calls
for a reliable and fast method to obtain matrix elements for 
all components of the TM
force in higher partial waves.

Recently we have proposed a novel, automatized way to perform a partial 
wave decomposition
of any two- and three-nucleon potential~\cite{Golak.2010}. 
This approach makes use of a software  for symbolic calculations to generate 
the part of the code which is specific for a considered force model.
More precisely, in this way we calculate exactly the isospin and 
spin-momentum parts of the nuclear interactions and generate a 
corresponding FORTRAN (or C) code.
That momentum dependent output forms an integrand for further
five-dimensional numerical integrations.

In this paper we present results of applying that new scheme 
to the original TM 3N force. They confirm the 
feasibility and efficiency of our method and its 
numerical implementation. 
The existence of such a reliable procedure is especially important in 
view of available and forthcoming results from the chiral perturbation
theory ($\chi$PT) \cite{Bernard}
for 3N forces at higher orders of the chiral expansion. 
 A big number of different momentum-spin-isospin structures
 contained in those interactions 
 requires a safe and automatized method to perform partial wave decompositions,
which is guaranteed by our method.
Furthermore our scheme avoids the application of partial wave decomposed 
permutation operators when dealing with products of 3NF's and permutation 
operators as they are often required, e.g., in 3N Faddeev equations. 
Such an application is numerically demanding because it requires a huge 
number of partial waves. 
Thus, again an efficient, fast and precise method is needed.

Our novel scheme of an automatized partial wave decomposition (aPWD) 
is described in 
Sec.~\ref{section2}. 
Results and additional tests for our numerical realization 
are presented in Sec.~\ref{section3} and conclusions are given in
Section~\ref{conclusion}.
The standard 
PWD of the \pirho component of the TM 3NF is given in the Appendix.

\section{Automatized partial wave decomposition}
\label{section2}

The 3NF, $V_{123}$, is an indispensable ingredient in a
theoretical description of the few-body systems. 
It can be always written as a sum of three terms
\begin{eqnarray}
V_{123} = V^{(1)} +  V^{(2)} +  V^{(3)} ~,
\label{e3nf_split}
\end{eqnarray}
where each $V^{(i)}$ is symmetrical under the exchange of nucleons $j$ and $k$
  ($i, j, k = 1, 2, 3$, $i \ne j \ne k$). 
Such a splitting 
in the case of the $\pi-\pi$ exchange TM 3NF 
corresponds to the possible choices of the nucleon undergoing
off-shell $\pi$N scattering. 

The 3NF typically
enters the dynamical equations via its part $V^{(1)}$.
In the case of the three-nucleon bound state, 
the Faddeev component $\psi$ fulfils the following 
equation~\cite{Nogga.1997}
\begin{equation}
\psi = G_0 t P \psi + (1+ G_0 t) G_0 V^{(1)} (1+P) \psi \;,
\label{eq.bs}
\end{equation}
where $G_0$ is the free 3N propagator and $t$ is the two-body
$t$-operator generated from a given nucleon-nucleon (NN) potential through the
Lippmann-Schwinger equation. 
The permutation operator $P \equiv P_{12}P_{23}+P_{13}P_{23}$ is given 
in terms of the transpositions $P_{ij}$,
which interchange particles $i$ and $j$. 
The full bound state wave function $ \Psi $ is then obtained 
as $ \Psi = ( 1 + P ) \psi $.

Transition amplitudes for the elastic
nucleon-deuteron scattering, $U$, and for the breakup reaction, $U_0$,  
are given as~\cite{HuberAPP}
\begin{eqnarray}
U&=&PG_0^{-1}\Phi + PT + V^{(1)} (1+P)\Phi + V^{(1)} (1+P)G_0T \;,\cr
U_0 &=& (1+P)T\;,
\end{eqnarray}
where the  auxiliary state $T$ fulfils the 3N Faddeev equation
\begin{equation}
T=tP\Phi + (1+t G_0) V^{(1)} (1+P) \Phi + t P G_0 T 
+ (1+t G_0) V^{(1)} (1+P)G_0 T \,,
\label{eq.cont}
\end{equation}
with $\Phi$ being the initial state composed of the deuteron wave function
and a momentum eigenstate of the projectile nucleon.

Equations (\ref{eq.bs}) and (\ref{eq.cont}) are 
solved \cite{HuberAPP,physrep_96} in 
the momentum space using 
 3N partial-wave states $\mid p, q, \alpha \rangle$ in 
the $jJ$-coupling~ \cite{book,physrep_96} 
\begin{eqnarray}
\mid p, q, \alpha \rangle \equiv
\mid  p q (l s ) j (\lambda \frac12 ) I (j I ) J M_J \rangle \mid (t \frac12 )
T M_T \rangle  \ ,
\label{eqn.alpha}
\end{eqnarray}
where $p$ and $q$ are magnitudes of the standard Jacobi momenta and 
$\alpha$ denotes a set of discrete quantum numbers arising in the following way:
the spin $s$ of the subsystem composed from nucleons 2 and 3 
is coupled with their orbital angular momentum $l$ 
to the total angular momentum $j$. The spin $\frac12$ 
of the spectator particle $1$
couples with its relative orbital angular momentum $\lambda$ 
to the total angular momentum of nucleon $1$, $I$.
Finally, $j$ and $I$ are coupled to the total 3N
angular momentum $J$ with the projection $M_J$. For the isospin part, 
 the total isospin t of the  $(23)$ subsystem is coupled with 
the isospin $\frac12$ of
the spectator nucleon to the total 3N isospin $T$ with the projection $M_T$.

Any three-nucleon force enters 
 Eqs.~(\ref{eq.bs})-(\ref{eq.cont}) in the form of 
 $V^{(1)}(1+P)$. Therefore
 a partial wave decomposition of $V^{(1)}$ as well as $V^{(1)}P$ 
 has to be performed.
 The standard approach to perform 
 a partial wave decomposition of $V^{(1)}$~\cite{Huber.1994} is 
 very tedious, even with improvements suggested in~\cite{Lazauskas}, 
since each momentum-spin-isospin structure, which occurs in a 3NF,
has to be treated separately. 
In the case when a 3NF consists of a big number of such structures, like 
chiral 3NF's at 
higher orders of the chiral expansion, the traditional
approach to a
partial wave decomposition is very inefficient and extremely time consuming.
 In addition, the application of the permutation operator, when calculating 
$V^{(1)}P$,  causes an additional  numerical problem,
which originates from  a slow convergence of the $V^{(1)}P$ 
matrix elements 
with respect to the number of intermediate states $|\alpha''>$:
\begin{eqnarray}
&&\langle p,q,\alpha \mid V^{(1)}P \mid p',q',\alpha' \rangle = \\ \nonumber
&=& \int dp'' p''^2 \int dq'' q''^2 \sum_{\alpha''}
\langle p,q,\alpha \mid V^{(1)} \mid p'',q'',\alpha'' \rangle 
\langle p'',q'',\alpha'' \mid P \mid p',q',\alpha' \rangle \;.
\label{eq6}
\end{eqnarray}
In order to calculate precisely these matrix elements,
a big number of intermediate states $\mid \alpha'' \rangle$ is required, 
and, thus, one is forced to calculate matrix elements of $V^{(1)}$ operator for
a much bigger set of $\alpha''$ states than 
actually needed 
in order to get converged 
solutions of the Faddeev equations.

In our new approach, called in the following
automatized partial wave decomposition (aPWD),
 to get matrix elements of  $V^{(1)}$ and
$V^{(1)}P$, that drawback is
 removed because matrix elements of $V^{(1})$ and $V^{(1)}P$ are
 calculated directly.

The starting point of our method is the
observation, that 
any 3N interaction and thus also 
its $V^{(1)}$ component in momentum space 
can be written as a sum of 
terms in the form
\begin{equation}
V^{(1)} = f(\vec{q_1},\vec{q_2},\vec{q_3}) 
\hat O_{spin}(\vec{q_1},\vec{q_2},\vec{q_3},\vec{\sigma_1},\vec{\sigma_2},\vec{\sigma_3})
 \hat O_{isospin}(\vec{\tau_1}, \vec{\tau_2}, \vec{\tau_3})
\end{equation}
where $\hat O_{spin}$ and $\hat O_{isospin}$ are operators acting 
on spin and isospin degrees of freedom, respectively, which are built from 
the spin ($\vec \sigma_i$) and isospin ($\vec \tau_i$) operators of 
individual nucleons.
Scalar factors $f(\vec{q_1},\vec{q_2},\vec{q_3})$ and  spin operators
$ \hat O_{spin}(\vec{q_1},\vec{q_2},\vec{q_3},\vec{\sigma_1},\vec{\sigma_2},\vec{\sigma_3)}$
depend on momentum transfers $\vec{q_i}$ to nucleon $i$ which are 
 expressed in terms of 
the initial and final Jacobi momenta $\vec{p},\vec{q}$ and $\vec{p'},\vec{q'}$, respectively,
as:
\begin{equation}
\vec{q_1}=\vec{q'}-\vec{q}\; , \;\;\;\;\vec{q_2}
=(\vec{p'}-\vec{p})-\frac12(\vec{q'}-\vec{q})\; , \;\;\;\;
\vec{q_3}=-(\vec{p'}-\vec{p})-\frac12(\vec{q'}-\vec{q})=-(\vec{q_1}+\vec{q_2})\,.
\end{equation}

For example, in the \pipi part of the TM 3N force one meets the
following spin-isospin structures:
\begin{eqnarray}
\hat O_{spin}(\vec{q_1},\vec{q_2},\vec{q_3},\vec{\sigma_1},
\vec{\sigma_2},\vec{\sigma_3})&=& (\vec{\sigma_2}\cdot\vec{q_2}) \; 
(\vec{\sigma_3}\cdot\vec{q_3})
 \;,\;\;\;\; 
(\vec{\sigma_2}\cdot\vec{q_2}) \; (\vec{\sigma_3}\cdot\vec{q_3})\; 
(\vec{q_2}\cdot\vec{q_3}) \;,\;\; \nonumber \\
&&(\vec{\sigma_2}\cdot\vec{q_2}) \; (\vec{\sigma_3}\cdot\vec{q_3}) 
\; ((\vec{q_2})^2+(\vec{q_3})^2)  \;,\;\;\;\; 
\vec{\sigma_1}\cdot \vec{q_2} \times \vec{q_3} \\ \nonumber
\hat O_{isospin}(\vec{\tau_1},\vec{\tau_2},
\vec{\tau_3})&=& \vec{\tau_2} \cdot \vec{\tau_3} \;,\;\;\;\; 
i\vec{\tau_1} \cdot \vec{\tau_2} \times \vec{\tau_3} \;.\nonumber
\end{eqnarray}
Note, that not all combinations of $\hat O_{spin}$ and 
$\hat O_{isospin}$ actually appear in the above  example.

In the first step of aPWD we calculate 3NF matrix elements using
partial wave states $\mid p, q, \beta \rangle$ \cite{book}
in the so-called $LS$-coupling
\begin{eqnarray}
\mid p, q, \beta \rangle_1 \equiv
\mid p q (l \lambda ) L (s \frac12 ) S  (L S ) J M_J \rangle_1 
\mid (t \frac12 ) T M_T \rangle_1  \;,
\label{pqbeta}
\end{eqnarray}
where  
the relative orbital angular momentum $l$
(within the pair $(23)$) and $\lambda$ (between the pair $(23)$ and nucleon $1$)
are coupled to the total  orbital angular momentum $L$.
In the spin space, the spin
of the  $(23)$ pair is coupled with the spin $\frac12$ of the nucleon $1$ to the
total spin $S$. Finally $L$ and $S$ are coupled to the total 3N
angular momentum $J$ with the projection $M_J$.
The index 1 emphasizes, that the spectator particle is nucleon 1. 
$\beta$ describes the set of discrete quantum numbers discussed above.
The isospin state is the same as in the basis state $\vert p,q,\alpha\rangle$.

In this basis, it is easy to decouple the isospin and spin parts from
the momentum part, what leads to the following form of a 3NF matrix element
\begin{eqnarray}
&& 
\langle p' q' (l' \lambda' ) L' (s' \frac12 ) S'  (L' S' ) J M_J 
\mid \langle (t' \frac12 ) T' m_{T'} \mid
V^{(1)} 
\mid  p q (l \lambda ) L (s \frac12 ) S  (L S ) J M_J  \rangle 
\mid (t \frac12 ) T M_T \rangle 
\nonumber \\
& =& \int d \hat{p\,}' \! \! \int d \hat{q\,}' \! \! 
\int d \hat {p} \! \! \int d \hat {q}  
\sum\limits_{m_{L'}} C ( L',S',J;m_{L'}, M_J - m_{L'},M_J) 
\sum\limits_{m_L} C ( L,S,J;m_L, M_J - m_L,M_J)
\nonumber \\
& \times &  
{\cal Y}_{l',\lambda'}^{*\, L' ,m_{L'}} ({\hat p}',{\hat q}' )  \,
{\cal Y}_{l ,\lambda }^{ L  ,m_{L }} ({\hat p} ,{\hat q}  )  \,
\langle (s' \frac12 ) S' M_J -m_{L'}  \mid 
\hat O_{spin} (\vec{p'},\vec{q'},\vec{p},\vec{q}) 
\mid (s \frac12) S  \,M_J - m_L \rangle  \nonumber \\
&\times& f(\vec{p'},\vec{q'},\vec{p},\vec{q}) 
\langle (t' \frac12 ) T' M_T  \mid
\hat O_{isospin} \mid (t \frac12) T M_T\rangle, 
\label{PWD11}
\end{eqnarray}
where
\begin{equation}
{\cal Y}_{l, \lambda}^{L, m_L} ( {\hat p} ,{\hat q} )
\equiv 
\sum\limits_{m_l=-l}^{l} C( l, \lambda, L ; m_l, m_L - m_l , m_L ) 
\, Y_{l,m_l} ( {\hat p} ) \,
Y_{\lambda,m_L-m_l} ( {\hat q} )\;. 
\label{calY}
\end{equation}
with the standard Clebsch-Gordan coefficients and the spherical harmonics.
For abbreviation we skip in (\ref{PWD11}) and in the following 
the spin $\vec{\sigma_i}$ and the isospin $\vec{\tau_i}$ 
operators in the arguments of $\hat O_{spin}$ and $\hat O_{isospin}$ operators.  

The matrix element in the spin space appearing in (\ref{PWD11}),
\mbox{$\langle (s' \frac12 ) S' M_J -m_{L'}  \mid 
\hat O_{spin} (\vec{p'},\vec{q'},\vec{p},\vec{q}) 
\mid (s \frac12) S  \,M_J - m_L \rangle$}, 
 depends on the 
 momenta $\vec q_i$ and spin quantum numbers.
Using a software for symbolic calculations 
(such as Mathematica$^\copyright$~\cite{math}
in our case) it is
very easy to calculate this matrix element
for all combinations of spin quantum numbers
as a function of the momentum vectors $\vec q_i$.
To this aim
we use the Kronecker product built in Mathematica,
which allows us to express the spin matrix element in terms
of simple matrix operations. This is even more straightforward in the case
of the isospin matrix element, which does not depend on any additional parameters.
Another advantage of using software for symbolic calculations is
the possibility to generate a Fortran (or C) code in an automatized way.
This eliminates possible errors which can be introduced during
programing of very lengthy formulas for the spin matrix element.
The calculation of the 3NF matrix elements requires finally
an eight-dimensional
integration shown in~(\ref{PWD11}). 
In a typical case the total isospin and its projection is conserved.
We also assume that the considered 3N force is rotationally invariant.
Then the matrix elements in~(\ref{PWD11}) vanish unless $J=J'$ and $M_J=M_{J'}$, and,
additionally, do not depend on $M_J$. Thus we can calculate 
\begin{eqnarray}
& & G ( l',\lambda',L',s',S',t',l,\lambda,L,s,S,J,t,T,M_T ) \equiv \frac{1}{2 J +1} \cr
&  \times & \sum\limits_{M_J=-J}^J \,
\langle (t' \frac12) T,M_T\mid \;
\langle p' q' (l' \lambda' ) L' (s' \frac12 ) S'  (L' S' ) J M_J \mid 
V^{(1)} \mid
 p q (l \lambda ) L (s \frac12 ) S  (L S ) J M_J \rangle \; \nonumber \\
&\times&\mid (t \frac12) T,M_T \rangle \,, 
\label{G}
\end{eqnarray}
which is equal to the original matrix element of $V^{(1)}$ 
given in Eq.(\ref{PWD11}).
The integrand in $G( l',\lambda',L',s',S',t',l,\lambda,L,s,S,J,t,T,M_T )$:
\begin{eqnarray}
&& \int d \hat{p\,}' \! \! \int d \hat{q\,}' \! \! \int d \hat {p} 
\! \! \int d \hat {q}
\frac{1}{2 J +1}
\sum\limits_{M_J=-J}^J
\sum\limits_{m_{L'}} C ( L',S',J;m_{L'}, M_J - m_{L'},M_J) \nonumber \\
& \times &
\sum\limits_{m_L} C ( L,S,J;m_L, M_J - m_L,M_J)
{\cal Y}_{l',\lambda'}^{*\, L' ,m_{L'}} ({\hat p}',{\hat q}' )  \,
{\cal Y}_{l ,\lambda }^{ L  ,m_{L }} ({\hat p} ,{\hat q}  ) \nonumber \\ \,
& \times &
\langle (s' \frac12 ) S' M_J -m_{L'}  \mid
\hat O_{spin} (\vec{p'},\vec{q'},\vec{p},\vec{q}) \mid (s \frac12) 
S  \,M_J - m_L \rangle  \nonumber \\
&\times& f(\vec{p'},\vec{q'},\vec{p},\vec{q}) \langle (t' 
\frac12 ) T' M_T  \mid
\hat O_{isospin} \mid (t \frac12) T M_T\rangle
\label{eq.G}
\end{eqnarray}
is a scalar and thus does not depend on all directions of the Jacobi
momenta~\cite{Balian_Berezin}.
Therefore we are free to choose for example $\vec p$ along z-axis 
($\vec{p}=(0,0,p)$) and $\phi_q=0$ and are left
with five-fold integrations only
\begin{eqnarray}
&&G ( l',\lambda',L',s',S',t',l,\lambda,L,s,S,J,t,T,M_T ) \cr
&=& 8 \pi^2 \int d \hat{p\,}' \! \! \int d \hat{q\,}' \; \! \! 
\! \! \int dcos(\theta_q) \; \! \!
\frac{1}{2 J +1}
\sum\limits_{M_J=-J}^J
\sum\limits_{m_{L'}} C ( L',S',J;m_{L'}, M_J - m_{L'},M_J) \nonumber \\
& \times &
\sum\limits_{m_L} C ( L,S,J;m_L, M_J - m_L,M_J)
{\cal Y}_{l',\lambda'}^{*\, L' ,m_{L'}} ({\hat p}',{\hat q}' )  \,
{\cal Y}_{l ,\lambda }^{ L  ,m_{L }} ({\hat z} ,{\hat q}=(\sin(\theta_q),0,\cos(\theta_q))  ) \nonumber \\ \,
& \times &
\langle (s' \frac12 ) S' M_J -m_{L'}  \mid
\hat O_{spin} (\vec{p'},\vec{q'},\vec{p}=(0,0,p),\vec{q}=q(\sin(\theta_q),0,\cos(\theta_q))) 
\mid (s \frac12)
S  \,M_J - m_L \rangle  \nonumber \\
&\times& f(\vec{p'},\vec{q'},\vec{p}=(0,0,p),\vec{q}=q(\sin(\theta_q),0,\cos(\theta_q))) \langle (t'
\frac12 ) T M_T  \mid
\hat O_{isospin} \mid (t \frac12) T M_T\rangle
\label{eq.G5dim}
\end{eqnarray}
The reduction of the number of integrations for a simple example of 3NF
is numerically exemplified in Ref.~\cite{Golak.2010}.

The remaining summations over $m_{L'}, m_L$ and $M_J$ and 5-fold integrations 
can be performed for a small number of (p,q,p',q') combinations 
even on a personal computer.
However, a large number of
five-dimensional integrations, as they are needed to obtain all matrix 
elements needed for the solution of the 3N Faddeev equations,
has to be carried out on a powerful parallel computer.
Once the matrix elements 
\mbox{$\langle p',q', \beta' \mid V^{(1)} \mid p, q, \beta \rangle$}
are calculated,
recoupling to the $jI$-representation, 
\mbox{$\langle p',q', \alpha'\mid V^{(1)} \mid p, q, \alpha \rangle$,}
can be easily performed~\cite{book}
\begin{eqnarray}
&& \langle p',q', \alpha'\mid V^{(1)} \mid p, q, \alpha \rangle 
= \sum_{\beta,\beta'} \sqrt{(2j+1)\,(2J+1)\,(2L+1)\;(2S+1)} 
\left \{ {\begin{array}{ccc} l&s&j\\ \lambda&\frac{1}{2}&I \\ L&S&J \end{array}} \right \}
\nonumber\\
&\times&
\sqrt{(2j'+1)\,(2J'+1)\,(2L'+1)\;(2S'+1)} 
\left \{ {\begin{array}{ccc} l'&s'&j'\\ \lambda'&\frac{1}{2}&I' \\ L'&S'&J \end{array}} \right \}
\langle p',q', \beta' \mid V^{(1)} \mid p, q, \beta \rangle \;.
\end{eqnarray}

Now let us turn to the $V^{(1)}(1+P)$ 
operator and discuss its $V^{(1)}P_{12}P_{23}$ matrix element
\begin{eqnarray}
_1\langle p',q',\beta' \mid V^{(1)}P_{12}P_{23} \mid p,q,\beta \rangle_1 &=&
\int d\vec{\tilde{p'}} \int d\vec{\tilde{q'}}
\int d\vec{\tilde{p}} \int d\vec{\tilde{q}} \;\;
_1\langle p',q',\beta' \mid \vec{\tilde{p'}} \vec{\tilde{q'}} \rangle
\langle \vec{\tilde{p'}} \vec{\tilde{q'}} \mid V^{(1)} \nonumber \\
&\times&P_{12}P_{23} \mid \vec{\tilde{p}} \vec{\tilde{q}} \rangle
\langle \vec{\tilde{p}} \vec{\tilde{q}} \mid p,q,\beta \rangle_1 \,.
\end{eqnarray}
Since
\begin{eqnarray}
P_{12}P_{23} \mid \vec{\tilde{p}} \vec{\tilde{q}} \rangle_1 &=& 
\mid -\frac12 \vec{\tilde{p}} + \frac34 \vec{\tilde{q}}, 
-\vec{\tilde{p}} -\frac12 \vec{\tilde{q}} \rangle_1 P_{12}^{spin}P_{23}^{spin} 
\mid (s \frac12) S  \,M_S \rangle_1 \nonumber \\
&\times&P_{12}^{isospin}P_{23}^{isospin} \mid (t \frac12) T M_T\rangle_1\\
P_{12}^{spin}P_{23}^{spin} \mid (s \frac12) S M_S \rangle_1 
&=& \mid (s \frac12) S M_S \rangle_2 = \nonumber \\
&=& \sum_{s''} (-)^s \sqrt{\hat{s''} \hat{s} } 
{ \frac12 \quad \frac12 \quad s'' \brace \frac12 \quad S \quad s }  
\mid (s'' \frac12) S M_S \rangle_1 \\
P_{12}^{isospin}P_{23}^{isospin} \mid (t \frac12) T M_T \rangle_1 &=& 
\mid (t \frac12) T M_T \rangle_2 = \nonumber \\
&=& \sum_{t''} (-)^t \sqrt{\hat{t''} \hat{t} } { \frac12 \quad \frac12 
\quad t'' \brace \frac12 \quad T \quad t } 
\mid (t'' \frac12) T M_T \rangle_1\;,
\end{eqnarray}
where $P_{ij}^{spin} (P_{ij}^{isospin})$ is the part of $P_{ij}$ 
operator acting in the spin (isospin) space,
one gets
\begin{eqnarray}
&_1\langle p',q',\beta' & \mid V^{(1)}P_{12}P_{23} \mid p,q,\beta \rangle_1 =
\int d \hat{p\,}' \! \! \int d \hat{q\,}' \! \! \int d \hat {p} 
\! \! \int d \hat {q}
\sum\limits_{m_{L'}} C ( L',S',J;m_{L'}, M_J - m_{L'},M_J) \nonumber \\
&\times& \sum\limits_{m_L} C ( L,S,J;m_L, M_J - m_L,M_J)
\nonumber \\
& \times &
{\cal Y}_{l',\lambda'}^{*\, L' ,m_{L'}} ({\hat p}',{\hat q}' )  \,
{\cal Y}_{l ,\lambda }^{ L  ,m_{L }} ({\hat p} ,{\hat q}  )  \,
\sum_{s''} (-)^s  \sqrt{\hat{s''} \hat{s} } { \frac12 \quad \frac12 
\quad s'' \brace \frac12 \quad S \quad s }
\sum_{t''} (-)^t  \sqrt{\hat{t''} \hat{t} } { \frac12 \quad \frac12 
\quad t'' \brace \frac12 \quad T \quad t } \nonumber \\
& \times &
_1\langle (s' \frac12 ) S' M_J -m_{L'}  \mid
\hat O_{spin} (\vec{p'},\vec{q'},-\frac12 \vec{p} + \frac34 \vec{q},\;
-\vec{p}  -\frac12 \vec{q})
\mid (s'' \frac12) S  \,M_J - m_L \rangle_1 \nonumber \\
& \times & f(\vec{p'},\vec{q'},
-\frac12 \vec{p} + \frac34 \vec{q},
-\vec{p} -\frac12 \vec{q}) _1\langle (t' \frac12 ) T' M_T  \mid 
\hat O_{isospin} \mid (t'' \frac12) T M_T\rangle_1 \,.
\label{eq.p12p23}
\end{eqnarray}
Similarly, for $V^{(1)}P_{13}P_{23}$ one gets
\begin{eqnarray}
&_1\langle p',q',\beta' & \mid V^{(1)}P_{13}P_{23} \mid p,q,\beta \rangle_1 =
\int d \hat{p\,}' \! \! \int d \hat{q\,}' \! \! \int d \hat {p} \! \! 
\int d \hat {q}
\sum\limits_{m_{L'}} C ( L',S',J;m_{L'}, M_J - m_{L'},M_J) \nonumber \\
& \times & \sum\limits_{m_L} C ( L,S,J;m_L, M_J - m_L,M_J)
\nonumber \\
& \times &
{\cal Y}_{l',\lambda'}^{*\, L' ,m_{L'}} ({\hat p}',{\hat q}' )  \,
{\cal Y}_{l ,\lambda }^{ L  ,m_{L }} ({\hat p} ,{\hat q}  )  \,
\sum_{s''} (-)^{s''}  \sqrt{\hat{s''} \hat{s} } { \frac12 \quad \frac12 
\quad s'' \brace \frac12 \quad S \quad s }
\sum_{t''} (-)^{t''}  \sqrt{\hat{t''} \hat{t} } { \frac12 \quad \frac12 
\quad t'' \brace \frac12 \quad T \quad t } \nonumber \\
& \times &
_1\langle (s' \frac12 ) S' M_J -m_{L'}  \mid
\hat O_{spin} (\vec{p'},\vec{q'},-\frac12 \vec{p} - \frac34 \vec{q},\;
\vec{p} -\frac12 \vec{q})
\mid (s'' \frac12) S  \,M_J - m_L \rangle_1 \nonumber \\
& \times & f(\vec{p'},\vec{q'},
-\frac12 \vec{p} - \frac34 \vec{q},
\vec{p} -\frac12 \vec{q}) _1\langle (t' \frac12 ) T' M_T  \mid
\hat O_{isospin} \mid (t'' \frac12) T M_T\rangle_1.
\label{eq.p13p23}
\end{eqnarray}
That means that the calculation of these two contributions proceeds in the
same way as calculation of  the $V^{(1)}$ matrix element.  
Only the arguments of the term $\hat O_{spin}$
have to be changed and additional factors originating from 
the recoupling of the spin and isospin quantum  numbers 
have to be taken into account.
As for the $V^{(1)}$ operator also here the eight-fold 
integrations can be reduced to the
five-fold ones and recalculation to $\vert p,q,\alpha \rangle$ states can be 
performed.

It is important to note  that, 
 since our
basis states $\vert p,q,\alpha \rangle$ are antisymmetric with respect to the 
exchange of nucleons $2$ and $3$,  
 (2.19) and  (2.20) yield the 
same values for the matrix elements. 
This allows one  to reduce significantly the size of the codes and the
required computation time.

\section{Results}
\label{section3}

\subsection{The TM 3NF and its \pipi, \pirho, and \rhorho components}

Since the aim of this work is not to study the dependence of 
the matrix elements of the TM force on its parameters,
in the following we use their values given in Tab.I of Ref.~\cite{Coon.1993}:
 $a=1.03\mu^{-1}, b=-2.62\mu^{-3}, c=0.91\mu^{-3}, d=-0.753\mu^{-3}$ 
with $\mu=139.6$ MeV. 
In the numerical implementation of~(\ref{PWD11}) we use 
the same number of 
gaussian points for each of the five angular domains. 
It might be more efficient to relax this constraint in future applications and to optimize the grids further.
Thus our integration method leaves 
room for improvement, even if we will later demonstrate in subsection~\ref{subsecNINT}
that it leads to fully
converged results.

\begin{figure}[t]\centering
\epsfig{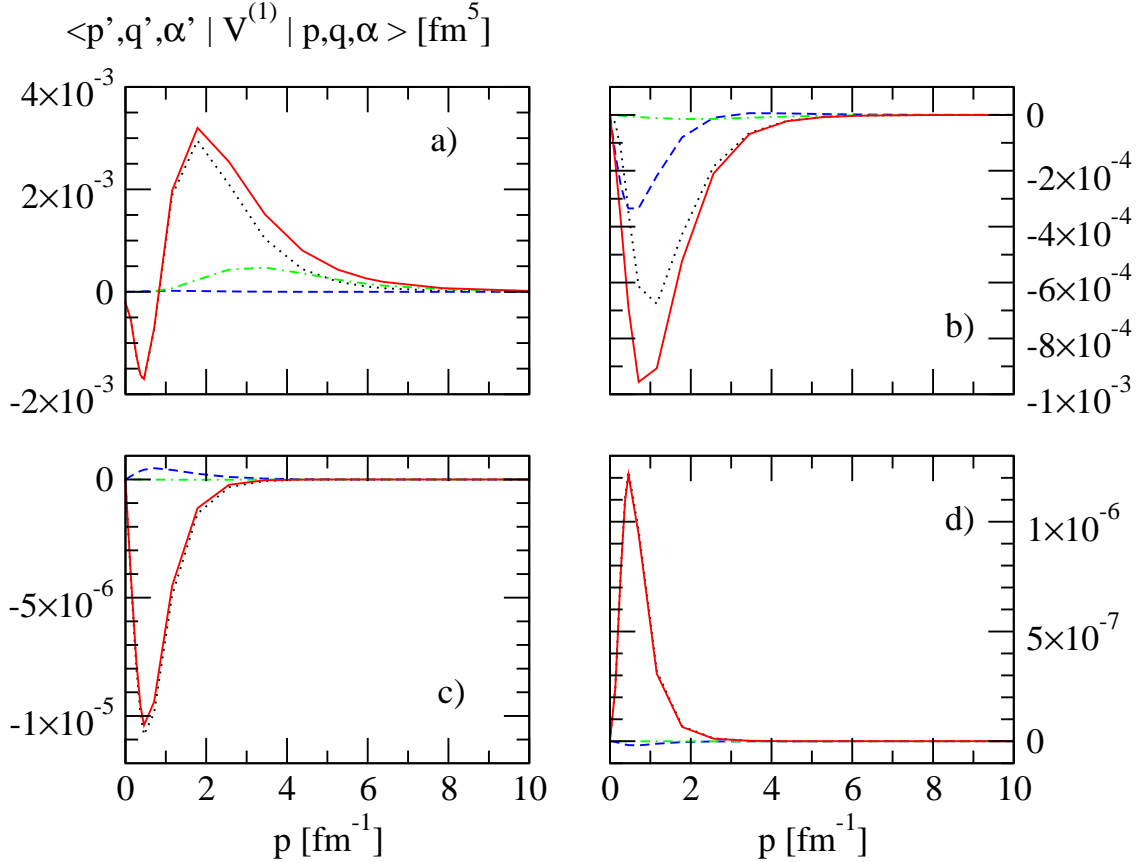}
\caption{(Color online) The TM 3NF matrix elements 
$< p'=0.132 fm^{-1}, q'=0.132 fm^{-1}, \alpha' \vert V^{(1)} \vert p,
  q=0.132 fm^{-1}, \alpha >$
as a function of the $p$ momentum
for ($\alpha', \alpha$): a) (1,1), b) (1,4), c) (6,3) d) (6,8).
The solid (red) curve represents the full TM 3NF and the other curves show the
contributions coming from the \pipi (black dotted), \pirho (blue dashed)
and \rhorho (green dot-dashed) components.}
\label{f1}
\end{figure}

\begin{figure}[t]\centering
\epsfig{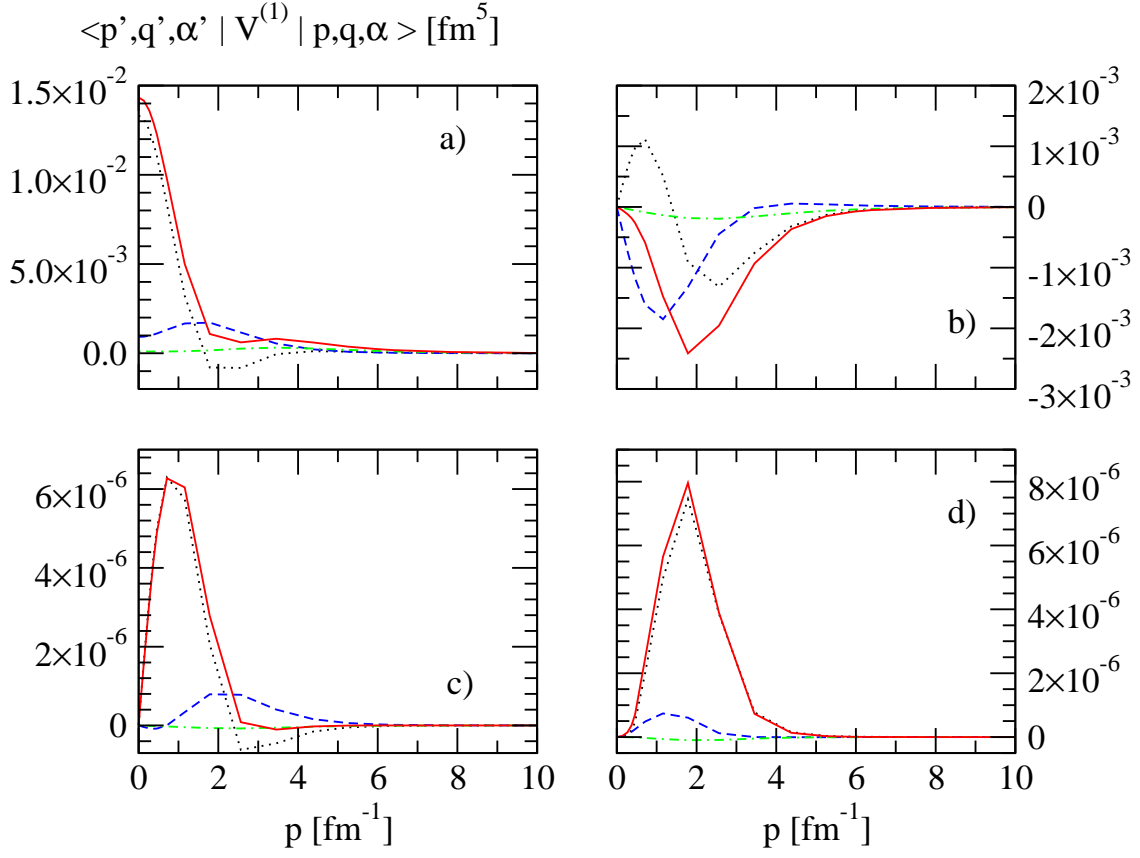}
\caption{(Color online) The same as in Fig.~\ref{f1} but
for momenta values: $p'=0.711 fm^{-1}, q'=0.132 fm^{-1}, 
q=2.842 \;{\rm fm}^{-1}$.}
\label{f2}
\end{figure}

The  TM 3NF matrix elements calculated in the basis (\ref{eqn.alpha})
are functions of four momentum magnitudes
and two sets of discrete quantum numbers.
In Figs.~\ref{f1}-\ref{f2}, examples of the TM force $V^{(1)}$ 
matrix elements
are shown together with its $\pi-\pi$,
$\pi-\rho$ and $\rho-\rho$
components in one-dimensional plots.
In Fig.~\ref{f1}, the matrix elements 
$< p', q', \alpha' \vert V^{(1)} \vert p, q, \alpha >$
for $p'=q'=q=0.132\; {\rm fm}^{-1}$ and for different channel 
pairs ($\alpha', \alpha$) (see Tab.~\ref{tab1}) are shown
as a function of the momentum p.
The same matrix elements but for the momenta
$p'=0.711$ fm$^{-1}$, $q'=0.132$ fm$^{-1}$, and $q=2.84$ fm$^{-1}$ 
are shown in Fig.~\ref{f2}
again as a function of $p$.
The $\pi-\pi$ part dominates in all cases but the $\pi-\rho$ 
part is also important (see Fig.~\ref{f1}b,\ref{f2}a-c). The 
$\rho-\rho$ part is of less importance 
for all the considered matrix elements.

\begin{table}
\begin{tabular}{|c|c|c|c|c|c|c|}
\hline
$\;\;\;\;\alpha$\;\;\;\; & \;\;\;\;l\;\;\;\; & \;\;\;\;s\;\;\;\; & \;\;\;\;j\;\;\;\; & \;\;\;\;$\lambda$\;\;\;\; & \;\;\;\;I\;\;\;\; & \;\;\;\;t\;\;\;\; \\
\hline
1 & 0 & 0 & 0 & 0 & $\frac12$ & 1 \\
3 & 1 & 0 & 1 & 1 & $\frac12$ & 0 \\
4 & 1 & 0 & 1 & 1 & $\frac32$ & 0 \\
6 & 0 & 1 & 1 & 2 & $\frac32$ & 0 \\ 
8 & 2 & 1 & 1 & 2 & $\frac32$ & 0 \\
\hline
\end{tabular}
\caption{The values of the discrete quantum numbers 
for selected $\alpha$ states (\ref{eqn.alpha}) for the total angular momentum $J=\frac12$ and the 
positive parity $\Pi=(-1)^{l+\lambda}$.}
\label{tab1}
\end{table}

\subsection{The aPWD for V$^{(1)}$(1+P) operator}

\begin{figure}[t]\centering
\epsfig{file=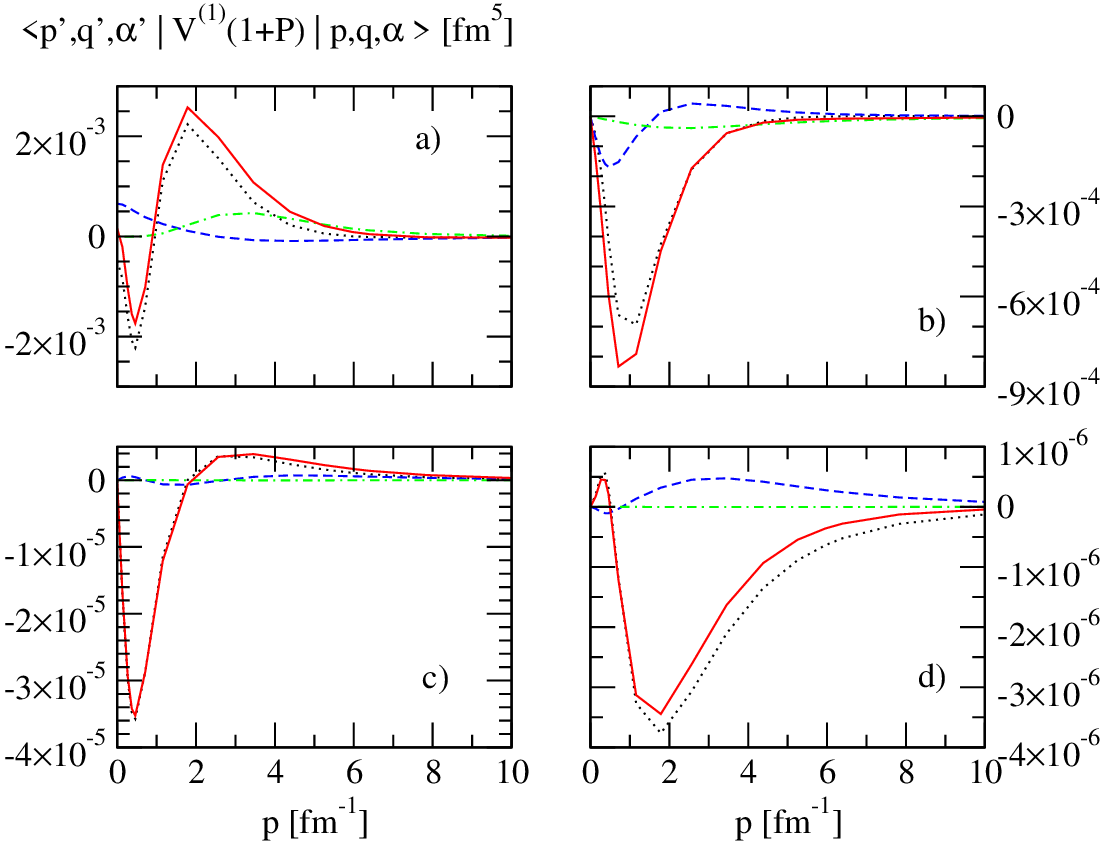,width=15cm,clip=true}
\caption{(Color online) The same as in Fig.~\ref{f1} but for the $V^{(1)}(1+P)$ operator.}
\label{f4}
\end{figure}

\begin{figure}[t]\centering
\epsfig{file=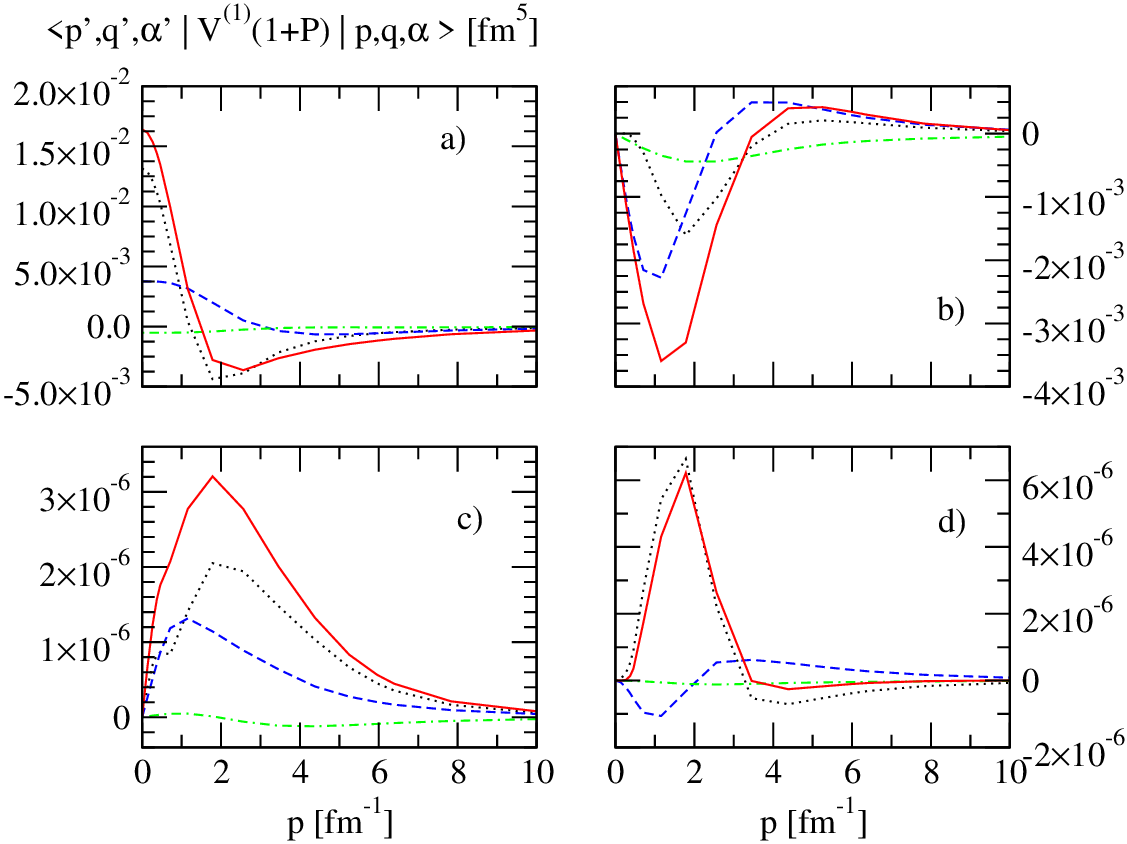,width=15cm,clip=true}
\caption{(Color online) The same as in Fig.~\ref{f2} but for the $V^{(1)}(1+P)$ operator.}
\label{f5}
\end{figure}

As was described in Sec.~\ref{section2}, aPWD can be applied 
not only to the $V^{(1)}$ alone but also to the $V^{(1)}(1+P)$ operator.
Using aPWD for $V^{(1)}(1+P)$ has the same advantages as for the 
$V^{(1)}$ operator: the automatized procedure 
can be easy tuned to any kind of 3NF
and 
reduces the possibility 
of errors. 
In the current implementation 
of aPWD the calculation of  
$V^{(1)}(1+P)$ matrix elements needs about the one and half amount of
the computing time needed for $V^{(1)}$, what is important from the practical 
point of view. Finally, in the standard scheme of PWD, the number of intermediate 
partial waves 
used to represent the P operator is limited and 
might be insufficient.
In the case of aPWD there is no separate decomposition of the
permutation operator what corresponds to the inclusion of all three-body intermediate waves. 
In Figs.~\ref{f4} and \ref{f5} the matrix elements of V$^{(1)}$(1+P)  
are shown for the same momenta and channels as in Figs.~\ref{f1} and
\ref{f2}, respectively.
For the channel combinations (1,1) and (6,3) in Fig.~\ref{f4}
and (1,1) and (6,8) in Fig.~\ref{f5}, where \pipi force dominates, 
the picture is 
similar to the corresponding ones in Figs.~\ref{f1} and \ref{f2}.
For the remaining channel combinations the differences 
are more visible, for example the
inclusion of the permutation operator for the \pipi component for the (6,8) pair 
in Fig.~\ref{f4}
leads to the change of the sign and strength of this force. 
In that case also the \pirho part becomes bigger after the permutation operator is applied.
Also for the (6,3) case in Figs.~\ref{f4} and \ref{f5} the action of 
the permutation operator changes the strength of the matrix element and increases the momentum range,
where both \pipi and \pirho components play a significant role.
For all the here presented cases the \rhorho force is much smaller than
the remaining interactions.

\begin{figure}[t]\centering
\epsfig{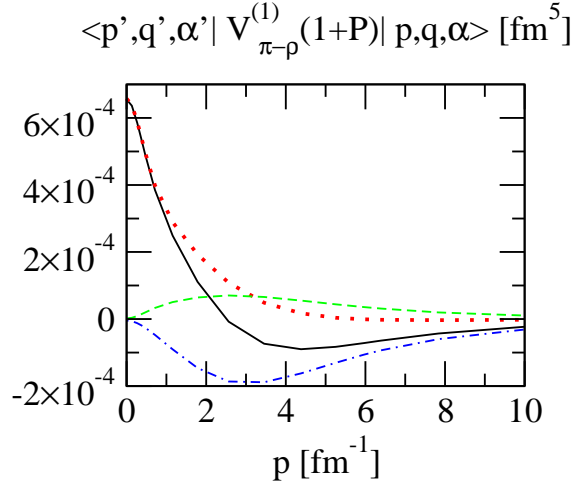}
\caption{(Color online) The contributions from the different parts of the \pirho force
for matrix elements
$< p'=0.132$ fm$^{-1}, q'=0.132$ fm$^{-1}, 
\alpha'=1 \vert V^{1}_{\pi-\rho}(1+P) \vert p, q=0.132$ fm$^{-1}, \alpha=1 >$.
The black solid line represents the total \pirho TM 3NF while
the red dotted, green dashed and blue dot-dashed lines represent the
"Kroll-Ruderman", the isospin even $\Delta$ and the isospin 
odd $\Delta$ terms, respectively.}
\label{f3}
\end{figure}

The aPWD method allows us to study the role played by different
isospin structures entering the TM force.
An example is given in Fig.~\ref{f3} where, for the \pirho force, the contribution
from the so-called "Kroll-Ruderman" and two "$\Delta$"
terms~\cite{Coon.1993} 
(see also Appendix \ref{krol-rud}) are shown.
For the presented matrix elements
($< p'=0.132 fm^{-1}, q'=0.132 fm^{-1}, \alpha'=1 \vert
V^{(1)}_{\pi-\rho}(1+P) \vert p, q=0.132 fm^{-1}, \alpha=1 >$)
the "Kroll-Ruderman" term dominates for small momenta p, while the two "$\Delta$" terms
are bigger for $p>2$ fm$^{-1}$. However, they have opposite signs, so their combined effect 
is weak and leads to a reduction of the strength of the dominant "Kroll-Ruderman" term.

\subsection{The comparison of the standard and automatized 
PWD schemes for \pipi and \pirho forces.}
\begin{figure}[t]\centering
\epsfig{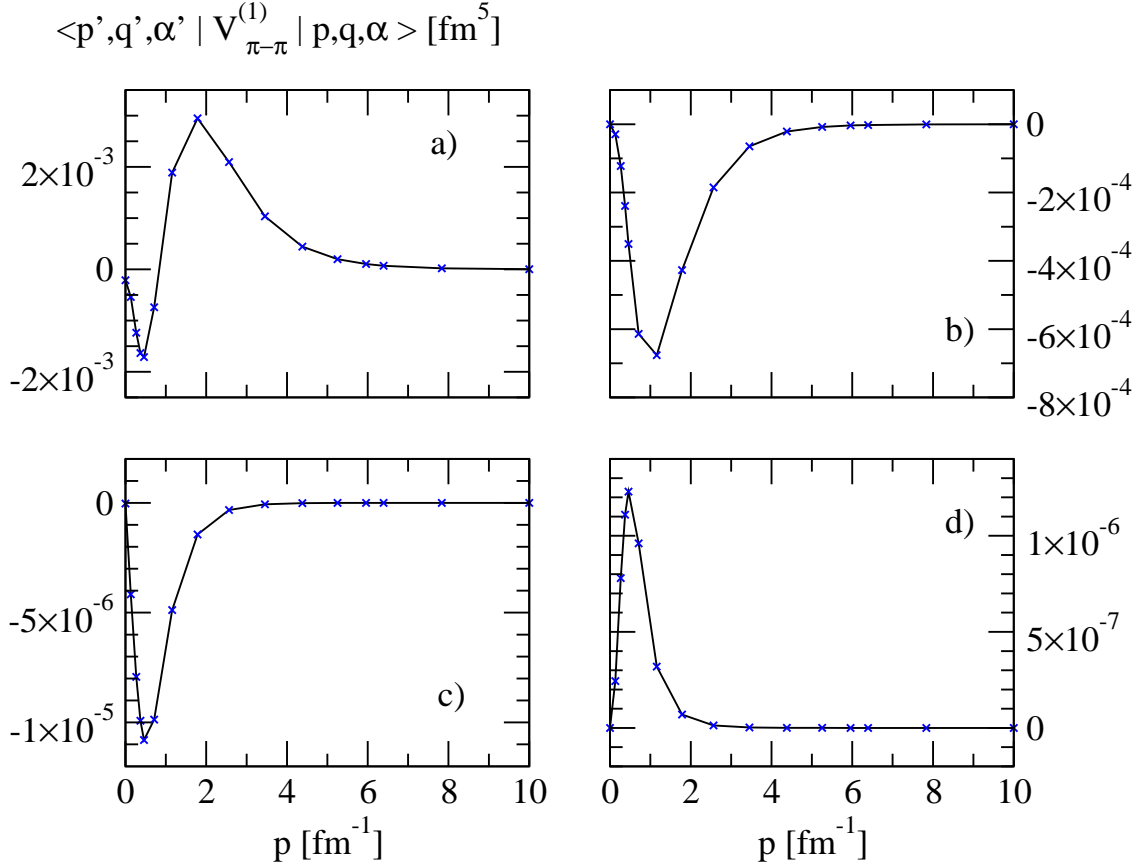}
\caption{(Color online) The comparison of the matrix elements of \pipi force
obtained in the standard (crosses) and automatized (solid line)
PWD. The channel combinations and momentum values are the same 
as in Fig.~\ref{f1}.
}
\label{f7}
\end{figure}

For the \pipi force the partial wave decomposition 
has been presented in ~\cite{CoonGlockle.1981} and
in an alternative way in ~\cite{Huber.1994}.
The comparison of results obtained by aPWD and the ones obtained in
Ref.~\cite{Huber.1994} is presented in Fig.~\ref{f7}.
Again the channel pairs and momenta are chosen as in Fig.~\ref{f1}.
A very good agreement between the both methods is clearly seen.

In Appendix A we present expressions for the partial wave decomposition of
the  \pirho force. This decomposition is in the spirit of the
decomposition of the \pipi interaction given in Ref.~\cite{Huber.1994}. 
In Fig.~\ref{f8} we compare results obtained in the aPWD scheme with those
based on PWD given in Appendix A.
Because of the internal construction of the PWD from Appendix A,
we compare matrix elements of $V^{(1)}P_{13}P_{23}$
instead of $V^{(1)}$.
The matrix elements of the standard PWD are obtained using partial waves 
up to $j_{max}=5$ in intermediate states. For this truncation, 
the matrix elements considered here are converged (see Sec.~\ref{subsec3e}). 
Again, for all given examples, the agreement between both methods is 
excellent.

\begin{figure}[t]\centering
\epsfig{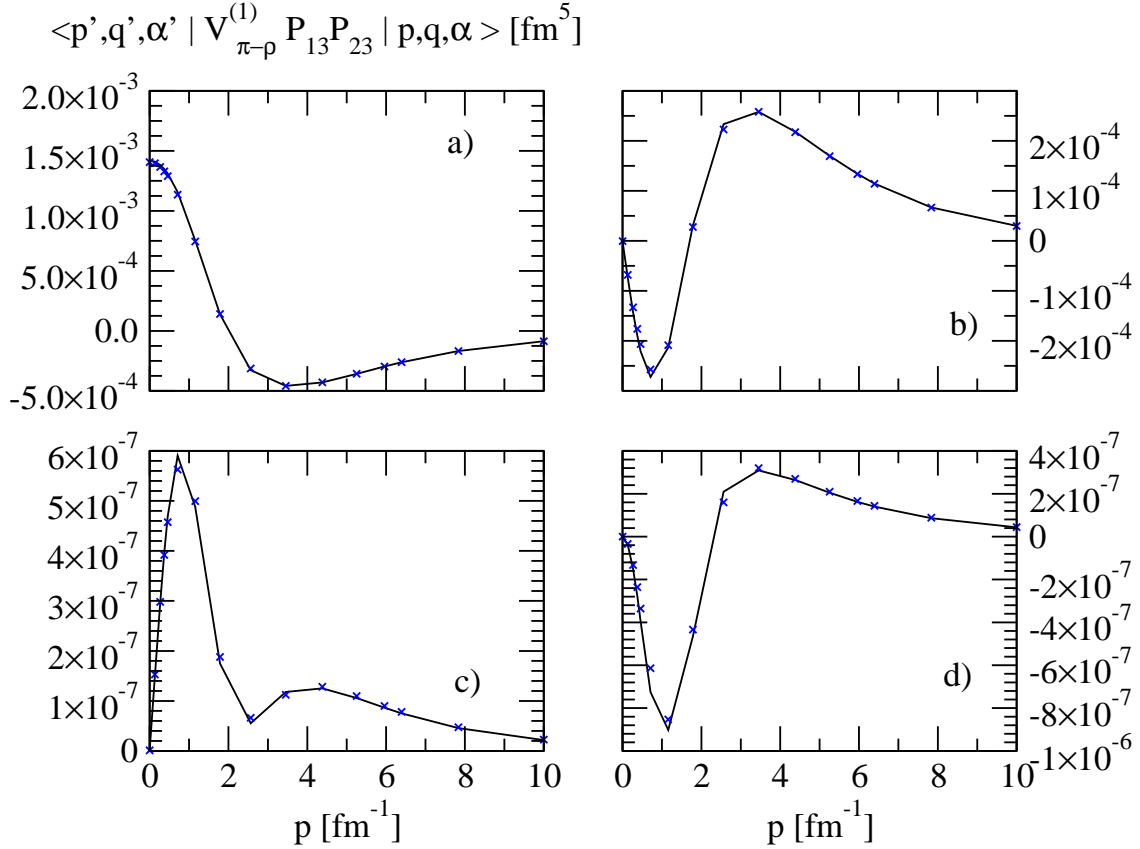}
\caption{(Color online) The comparison of the matrix elements of the \pirho force
obtained in the standard PWD from Appendix A (crosses) 
and automatized (solid line)
PWD. The channels pairs and momenta are the same as in Fig.~\ref{f2}.
}
\label{f8}
\end{figure}

Though the CPU time is smaller for the scheme presented in Appendix A,
the long time 
which is needed for the derivation of
the partial wave decomposition of complicated spin-momentum structures 
and its programming
in the standard way is incomparable with the relatively short 
time demanded by aPWD.
Another advantage of aPWD lies in its flexibility
which allows one to use it easily for different operators.
In the case of the standard PWD 
each spin-momentum structure has to be treated separately.

\subsection{The equality of $V^{(1)}P_{12}P_{23}$ and $V^{(1)}P_{13}P_{23}$}
\label{section4}

\begin{figure}[t]\centering
\epsfig{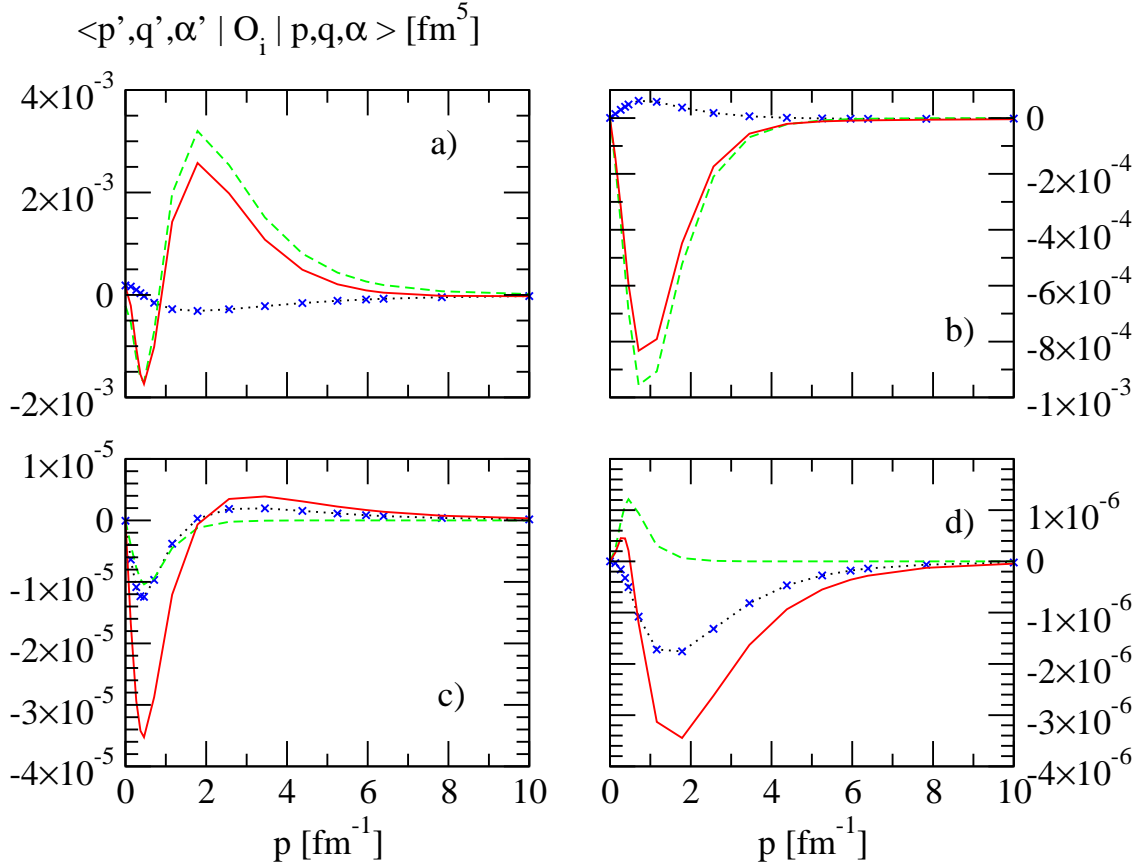}
\caption{(Color online) The contribution of $O_i \equiv
V^{(1)}$, $V^{(1)}P_{13}P_{23}$ and $V^{(1)}P_{12}P_{23}$ operators
to the total $V^{(1)}(1+P)$ TM 3NF matrix elements.
The channels combinations and momenta are chosen as in Fig.~\ref{f1}.
The crosses represent $V^{(1)}P_{13}P_{23}$ matrix elements.
The dashed, dotted and solid lines represent
$V^{(1)}$, $V^{(1)}P_{12}P_{23}$ and $V^{(1)}(1+P)$ matrix elements,
respectively.
}
\label{f6}
\end{figure}

The equality of $V^{(1)}P_{13}P_{23}$ and $V^{(1)}P_{12}P_{23}$ matrix elements
between the states antisymmetrized in the (23) subsystem
forms another nontrivial test of numerics.
To check this, we  
compare some matrix elements for $V^{(1)}P_{12}P_{23}$ obtained via Eq.(\ref{eq.p12p23})
with the corresponding ones for $V^{(1)}P_{13}P_{23}$ from Eq.(\ref{eq.p13p23}).
Results are displayed in Fig.\ref{f6} again for
four combinations of channel pairs and selected values of $p',q'$ and $q$ momenta (the same as in Fig.~\ref{f1}).
The numerical confirmation of the equality of the $V^{(1)}P_{13}P_{23}$ and $V^{(1)}P_{12}P_{23}$
matrix elements is clear. 
They differ from the
$V^{(1)}$ elements, as can be seen for some examples
in Fig.~\ref{f6}. All three possibilities are shown:
for the channel combinations (1,1) and (1,4) $V^{(1)}$ dominates, while
$V^{(1)}P_{12}P_{23}$ and $V^{(1)}P_{13}P_{23}$ are much smaller.
For the (6,3) combination each operator gives a similar
contribution to $V^{(1)}(1+P)$. For the (6,8) choice
and momenta around 2 fm$^{-1}$ the contribution from
$V^{(1)}$ is much smaller than the remaining two.

\subsection{The convergence of $V^{(1)}(1+P)$ matrix elements with respect to the number of 
the intermediate partial waves
for the \pirho and the full TM forces.}
\label{subsec3e}

\begin{figure}[t]\centering
\epsfig{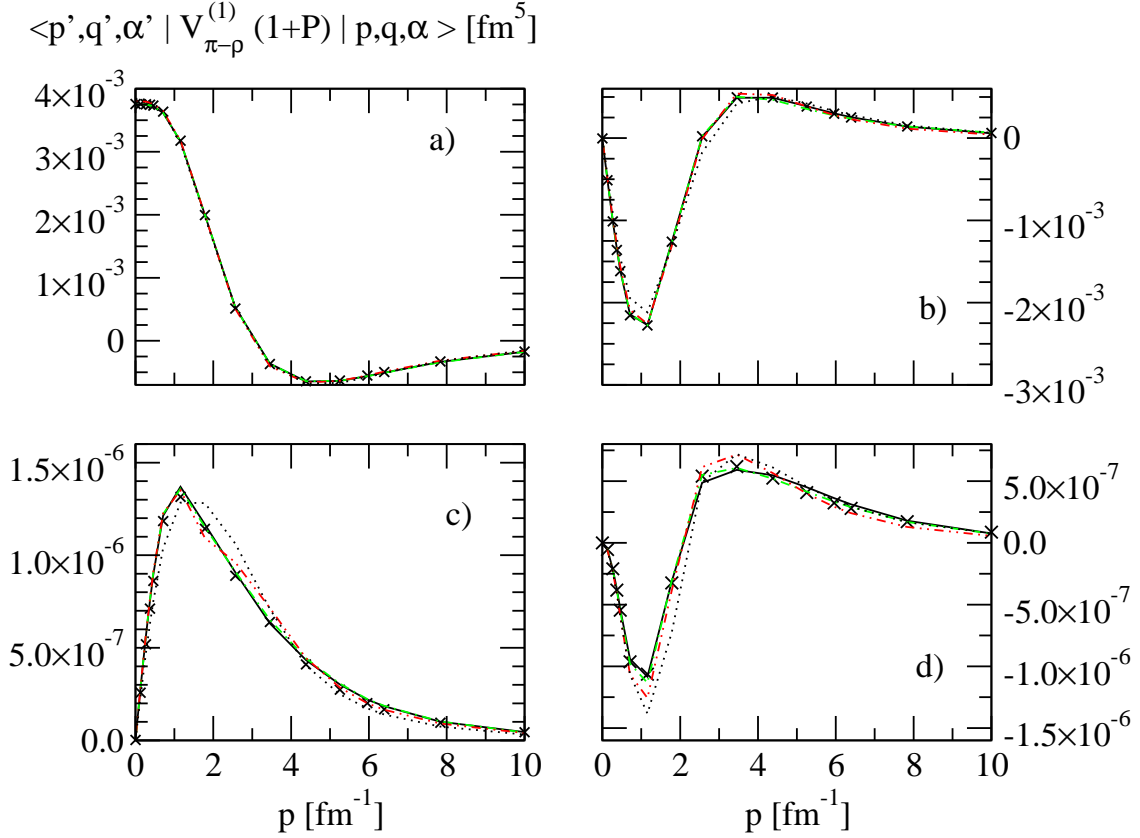}
\caption{(Color online) The convergence of the matrix elements of the \pirho part of the TM force: $V^{(1)}_{\pi-\rho}(1+P)$  
with respect to the number of the intermediate partial waves
used during the action of the permutation operator (see Eq.~(\ref{eq6})).
The channel combinations and momenta are the same as in Fig.~\ref{f2}. 
The crosses represent predictions obtained within the aPWD approach, the 
dotted (black), 
dash-double dotted (red), dash-dotted (green) and solid (black) lines 
represent results 
obtained with the traditional method described in Appendix A with all  
the intermediate 3N states up to $j_{max}=2,3,4$ and 5, respectively.
}
\label{f9}
\end{figure}

\begin{figure}[t]\centering
\epsfig{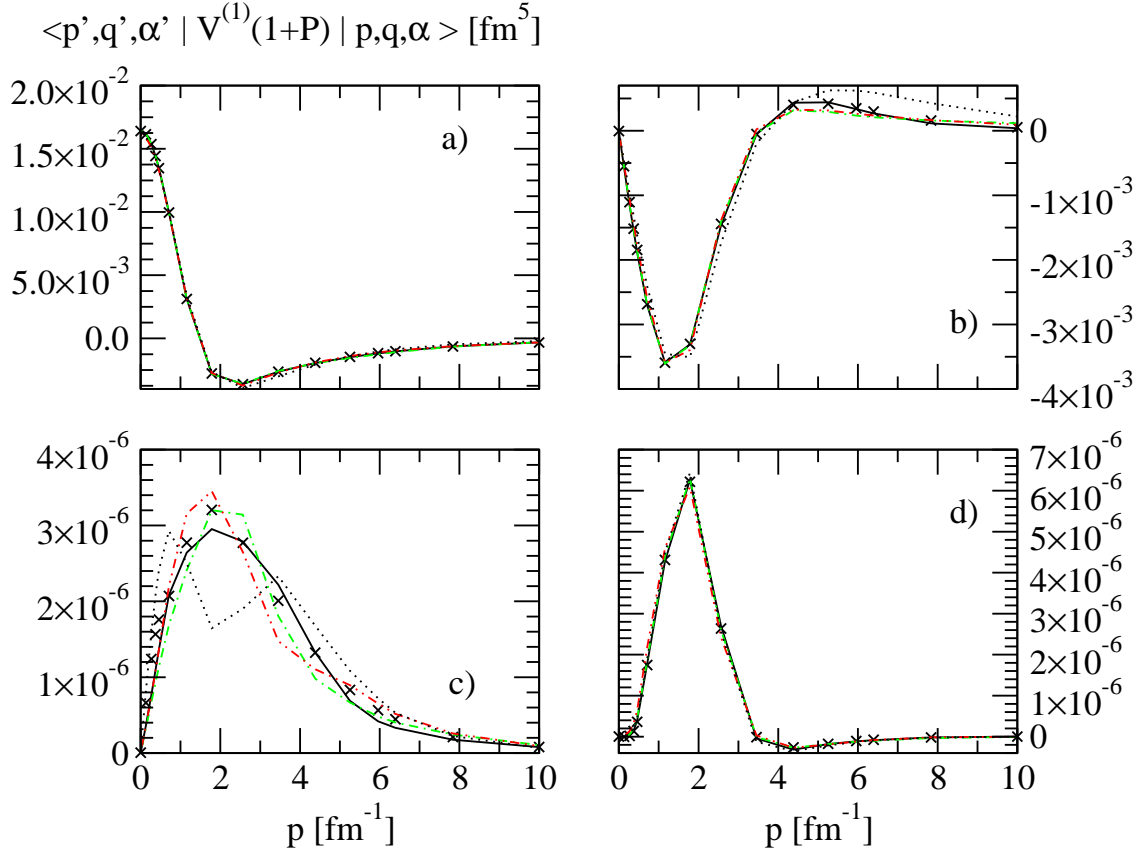}
\caption{(Color online)
The convergence of the full TM $V^{(1)}(1+P)$  matrix elements with
respect 
to the number of intermediate 
partial waves
used during the action of the permutation operator.
The channel combinations and momenta are as in Fig.~\ref{f2}.
The crosses represent predictions obtained within aPWD approach.
The dotted (black), dash-double dotted (red), dash-dotted (green) and 
solid (black) lines represent results 
obtained by the action of the permutation operator on the TM $V^{(1)}$ 
force with all 
the intermediate 3N states up to $j_{max}=2,3,4$ and 5, respectively.
}
\label{f10}
\end{figure}

The aPWD result for the $V^{(1)}(1+P)$ operator, which corresponds to the infinite number of the 
intermediate partial waves 
taken into account during the action of the permutation operator,
gives the limit to which results of the traditional scheme should converge.
This convergence is confirmed in Figs.~\ref{f9} and \ref{f10} for the
\pirho part of the TM and the full TM 3NF, respectively. The channels and momenta are 
the same as in 
Fig.~\ref{f2}. While for the channel combination (1,1) already the smallest number of
partial waves gives the aPWD limit, for the other combinations much
more partial waves have to be taken into account. For one of the shown here cases 
(Fig.~\ref{f10}c) taking all partial waves up to $j_{max}=5$ is still insufficient to achieve 
the limit of aPWD. Note, however, that the magnitude of this 
matrix element is relatively small.
In general, the convergence of the traditional PWD scheme is fully confirmed.

\subsection{The binding energy and correlation function for $^3$H.}
As a first application, we would like to calculate in the following
the binding
energy of $^3$H, some energy expectation values and the correlation function.
The obtained binding energies and expectation values of the kinetic energy 
$\langle H_0 \rangle$, the NN potential energy $\langle V_{NN} \rangle$ and 
the 3N potential energy $\langle V_{3N} \rangle$ 
 are given in Tab.~\ref{tab2}
for several realistic NN interactions alone and together with the TM force.
The TM force was included for all states with subsystem total angular momentum $j \leq 2$.
The inclusion of the TM force leads to a
stronger binding of $^3$H. The binding energy changes, after the inclusion of the TM force, by
approximately -1.093 MeV for the
CDBonn
potential and from -1.122 to -1.334 MeV for Nijmegen potentials.
These results are in a reasonable agreement with the ones given in Tab.~2 of
Ref.~\cite{Stadler.1995} for the Bonn OBEPQ (-9.596 MeV)  and the Nijmegen (-8.689 MeV) 
potentials. 
Note, that in Ref.~\cite{Stadler.1995} slightly different values of
the $a,b$ and $c$ parameters were used. 
In our calculations, we include partial waves up to $j_{max}=5$ for the two-body interaction. 
This is also different from Ref.~\cite{Stadler.1995} where only partial waves up to $j_{max}=2$ were included.
Of course, for the given set of parameters, the binding energies do not accurately reproduce the 
experimental value of -8.482 MeV. Because of the well-known scaling behavior of many N-d scattering 
observables with the triton binding energy (see for example~\cite{Witala_scaling}),
it will be necessary to finetune the TM model such that the triton binding
energy is more accurately reproduced, e.g. along the lines of Ref.~\cite{Nogga.1997}

The inclusion of the TM 3NF leads for all the NN potentials to higher expectation values of the kinetic energy and
lower expectation values of the NN potential energy (about 3-6 MeV).
The expectation values of the 3N potential energy amounts from 3.3\% to 4.8\% of
the expectation values of the NN potential, depending on the particular NN potential.
This observations are in line with the general expectations for the strength of 3NF's 
and the more compact state of $^3$H when the binding energy is increased.

\begin{table}
\begin{tabular}{|c|c|c|c|c|}
\hline
NN potential     & $E_t$ [MeV] & $\langle H_0 \rangle$ [MeV] & $\langle V_{NN} \rangle$ [MeV] & $\langle V_{3N} \rangle$ [MeV] \\
\hline
CDBonn           & -8.008 & 37.620 & -45.609 & - \\
Nijmegen I       & -7.738 & 40.737 & -48.467 & - \\
Nijmegen II      & -7.658 & 47.526 & -55.176 & - \\
Nijmegen 93      & -7.664 & 45.617 & -53.283 & - \\
\hline
CDBonn + TM      & -9.101 & 41.934 & -48.669 & -2.342 \\
Nijmegen I + TM  & -8.860 & 45.523 & -52.277 & -2.098 \\
Nijmegen II + TM & -8.992 & 54.318 & -61.112 & -2.189 \\
Nijmegen 93 + TM & -8.841 & 51.173 & -58.092 & -1.925 \\
\hline

\end{tabular}
\caption{The triton binding energies $E_t$ and the energy expectation values $\langle H_0 \rangle$, $\langle V_{NN} \rangle$ 
and $\langle V_{3N} \rangle$ for the different NN potentials alone and together with the TM 3NF.}
\label{tab2}
\end{table}

\begin{figure}[t]\centering
\epsfig{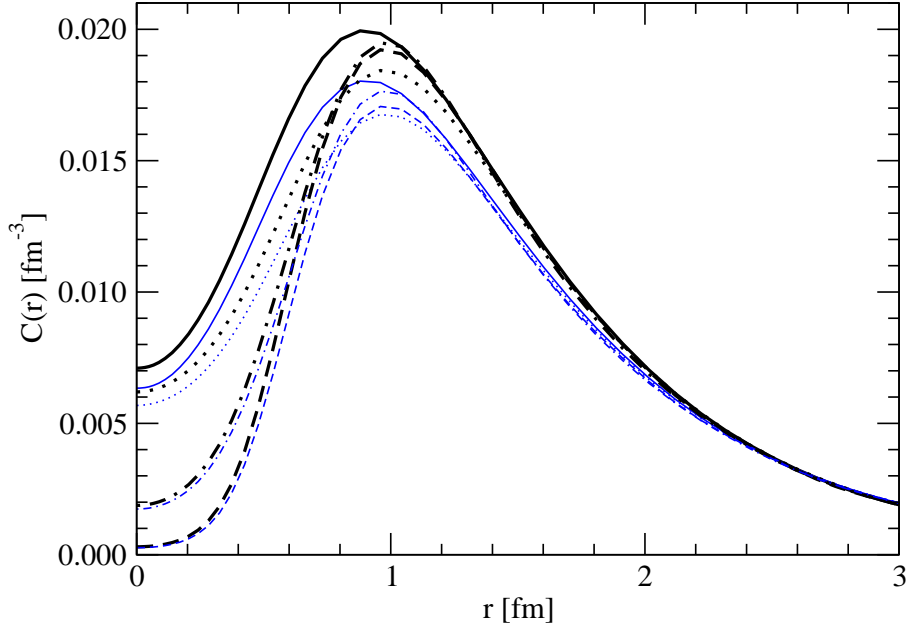}
\caption{(Color online) The two-body correlation function for the triton for different NN potentials alone (thin blue lines) and together with 
the TM 3NF (thick black lines).
Results obtained with the CDBonn, Nijmegen I, Nijmegen II and Nijmegen 93 potentials are represented by 
solid, dotted, dashed and dash-dotted curves, respectively.
}

\label{f11}
\end{figure}

The correlation function is defined in the configuration space as~\cite{Nogga.1997}
\begin{equation}
C(r) \equiv \frac13 \frac{1}{4\pi} \int d\hat{r} \langle \Psi \mid \sum_{i<j} \delta(\vec{r}-\vec{r_{ij}}) \mid \Psi \rangle
\end{equation}
where $r_{ij}$ is the
relative distance operator conjugate to the operator of the Jacobi momentum $\vec{p}$.
It is shown in Fig.~\ref{f11} for the different NN potentials alone and combined with the TM 3NF.
For the smaller distances shown in Fig.~\ref{f11}, the probability to find two nucleons increases when the 
TM 3NF is included. At least in part, this can be understood because the correlation functions drop more 
quickly for larger r due to the increased binding energy. Note that at short distances, 
the effect of the 3NF's is much larger than the dependence on the NN interaction model.
The here presented correlation functions are in good agreement with the ones presented in 
~\cite{Nogga.1997} for the same NN potentials combined with the \pipi part of the TM force.
 
\subsection{The quality of the five-dimensional integration}
\label{subsecNINT}

Finally, we would like to give an example of the stability of aPWD against the number
of points used in the numerical integrations. 
In Tab.~\ref{tab3} the
$V^{(1)}(1+P)$ matrix elements are given for the same channels and
momenta as in Fig.~\ref{f1} for two values of momentum p=0.711 ${\rm fm}^{-1}$  
and p=5.959 ${\rm fm}^{-1}$. Results were obtained using
N=12 or N=15 gaussian points in each of the five integrations in (\ref{PWD11}).
The agreement seen in Tab.~\ref{tab3} 
between both predictions 
clearly demonstrates 
that the numerical integration is well under control and 
leads to fully converged numbers.

\begin{table}
\begin{tabular}{|c|c|c|c|c|}
\hline
($\alpha',\alpha$) & N=12 & N=15 & N=12 & N=15 \\
& p=0.711 ${\rm fm}^{-1}$ & p=0.711 ${\rm fm}^{-1}$ & p=5.959${\rm fm}^{-1}$ & p=5.959${\rm fm}^{-1}$ \\
\hline
(1,1) & -0.0010139197  & -0.0010139197  & 8.9762335$\times 10^{-05}$ & 8.9762335$\times 10^{-05}$ \\
(1,4) & -0.00083291615 & -0.00083291615 & -9.0900123$\times 10^{-06}$ & -9.0900124$\times 10^{-06}$ \\
(6,3) & -2.8739479$\times 10^{-05}$ & -2.8739479$\times 10^{-05}$ & 1.6945111$\times 10^{-06}$ & 1.6945111$\times 10^{-06}$ \\
(6,8) & -1.212581$\times 10^{-06}$  & -1.21258$\times 10^{-06}$   & -3.5797036$\times 10^{-07}$ & -3.5797045$\times 10^{-07}$ \\
\hline
\end{tabular}
\caption{The $V^{(1)}(1+P)$ matrix elements (in ${\rm fm}^5$) for channels combinations and momenta $p',q',q$ 
as in Fig.~\ref{f1},
depending on the number of gaussian points N
used in the five-fold integration of Eq.(\ref{eq.G}).
The value of momentum $p$ is 0.711 ${\rm fm}^{-1}$ (left) and 5.959 ${\rm fm}^{-1}$ (right). }
\label{tab3}
\end{table}

\section{Summary}
\label{conclusion}

We apply an automatized method of partial wave decomposition to the Tucson-Melbourne
three-nucleon force. The obtained results agree very well with the traditional 
way of a partial wave decomposition for \pipi and \pirho contributions to the TM 3NF. 
For the latter one, we also give formulas of the partial wave decomposition in the traditional 
approach.
Matrix elements obtained in the new way are used in the calculations of the triton wave function 
with different underlying 
nucleon-nucleon potentials. We performed also different numerical tests, 
which confirm the reliability of our method and computer codes.

Among many advantages of aPWD, we would like to emphasize 
its generality, efficiency, the semi-automatized process of preparing a code and the
possibility of a calculation of the higher partial waves.
The latter point gives hope for the future use of the full Tucson-Melbourne
force in a description of 3N scattering at higher energies. The expected strong effects on observables 
coming from a 3NF should be tested also for short-range parts of three-body
interactions. Such parts are included in the Tucson-Melbourne force.

The automatized partial wave decomposition is especially  important 
in view of future applications 
 of 3NFs arising from the $\chi$PT. 
In this approach, consistent two- and three-body forces
are derived~\cite{Bernard}. The numerous spin-momentum and isospin
structures, which occur at higher orders of the 
chiral expansion require an efficient and automatized method 
for the PWD. The here presented results for the TM force prove, 
that such a method already exists. 

\section*{Acknowledgments}

This work was supported by 
the Polish Ministry of Science and Higher Education
under Grant No. N N202 077435. 
 It was also partially supported by the Helmholtz
Association through funds provided to the virtual institute ``Spin
and strong QCD''(VH-VI-231)  and by
  the European Community-Research Infrastructure
Integrating Activity
``Study of Strongly Interacting Matter'' (acronym HadronPhysics2,
Grant Agreement n. 227431)
under the Seventh Framework Programme of EU. 
 The numerical
calculations have been performed on the
 supercomputer cluster of the JSC, J\"ulich, Germany.

\appendix

\section{Standard PWD for \pirho component.}
\label{appendixa}
The $\pi$-$\rho$ part \cite{Stadler.1995} of the Tuscon-Melbourne 3NF
is given in terms of the momenta $\vec k_i$ and $\vec k_i~'$ of the individual
nucleons as 
\begin{eqnarray}
&&\langle \vec k_1' \vec k_2'\vec k_3'\vert W^{\pi\rho}_1 \vert \vec k_1 \vec k_2 \vec k_3 \rangle =
{ -1 \over (2\pi)^6 } { \delta(\vec k_1'+\vec k_2'+\vec k_3' -\vec k_1 -\vec k_2-\vec k_3)
\over (q^2+m_\rho^2)(q'^2+m_\pi^2)} (\sigma_3 \cdot q') 
\cr
&&
\times \{ -(i\vec \tau_1 \cdot \vec \tau_2 \times \vec  \tau_3)R^{\pi\rho}_{KR}(q^2,q'^2)
(i\vec \sigma_1 \cdot \vec \sigma_2  \times \vec q )  
\cr
&&
+(\vec \tau_2 \cdot \vec \tau_3 ) R^{\pi\rho}_{\Delta ^+} (q^2,q'^2) (\vec q\times \vec q~')\cdot 
(\vec q \times \vec \sigma_2) 
\cr
&&+(i\vec \tau_1 \cdot \vec \tau_2 \times \vec \tau_3)R^{\pi\rho}_{\Delta ^-} (q^2,q'^2) 
[(i\vec \sigma_1\cdot\vec \sigma_2\times \vec q~')q^2 - (i\vec \sigma_1\cdot \vec q\times \vec q~')
(\vec \sigma_2\cdot \vec q)] \} \cr
&&+ (2 \leftrightarrow 3, \vec q \leftrightarrow -\vec q~' ).
\end{eqnarray}
Introducing isospin, $I_+$ and $I_-$, and  spin, $F_{KR},
F_{\Delta+}^I, F_{\Delta+}^{II}, F_{\Delta-}^I,$ and $
F_{\Delta-}^{II} $, operators 
 we rewrite it as
\begin{eqnarray}
&&\langle \vec k_1'\vec k_2'\vec k_3'\vert W^{\pi\rho}_1 \vert \vec k_1 \vec k_2 \vec k_3 \rangle
\cr 
&&
=
{ -1 \over (2\pi)^6 }  \delta(\vec k_1'+\vec k_2'+\vec k_3' -\vec k_1 -\vec k_2-\vec k_3)
\cr
&&
\times \{ I_+ R^{\pi\rho}_{KR}(q^2,q'^2)
~F_{KR}
+I_- R^{\pi\rho}_{\Delta ^+} (q^2,q'^2)~ (F_{\Delta ^+}^I -F_{\Delta^+}^{II})
+I_+ R^{\pi\rho}_{\Delta ^-} (q^2,q'^2) ~
(F^I_{\Delta ^-}  +  F^{II}_{\Delta ^-}  )  \} \cr
&&+ (2 \leftrightarrow 3, q \leftrightarrow -q' ) 
\label{eq1}
\end{eqnarray}
with 
\begin{eqnarray}
I_+ \equiv i\vec \tau_1 \cdot \vec \tau_2 \times \vec \tau_3, ~~~
I_- \equiv \vec \tau_2 \cdot \vec \tau_3 ,
\label{Im}
\end{eqnarray}
\begin{eqnarray}
F_{KR} \equiv { -(\vec \sigma_ 3 \cdot \vec q~' ) \over { q'^2 + m_\pi ^2} }
{ i\vec \sigma_1 \cdot \vec \sigma_2 \times \vec q \over { q^2 + m_\rho ^2 }} ,
\label{Kroll-Ruderman}
\end{eqnarray}
\begin{eqnarray}
&&F_{\Delta ^+}\equiv { (\vec \sigma_ 3 \cdot \vec q~' ) \over { q'^2 + m_\pi^2} }
{(\vec q \times \vec q~') \cdot (\vec q\times \vec \sigma_2) \over  { q^2 + m_\rho ^2 }} 
=F_{\Delta^+}^I-F_{\Delta^+}^{II},
\cr && 
F_{\Delta^+}^I \equiv {(\vec \sigma_3 \cdot \vec q~')(\vec \sigma_2 \cdot \vec q~') \over 
{{q'}^2+m_\pi^2} } {q^2 \over q^2+ m_\rho^2},  ~~~
F_{\Delta^+}^{II}\equiv {(\vec \sigma_3 \cdot \vec q~') 
\over {q'}^2+m_\pi^2} (\vec q~' \cdot \vec q) {(\vec \sigma_2 \cdot \vec q) \over 
q^2+ m_\rho^2},
\end{eqnarray}
\begin{eqnarray}
F^I_{\Delta ^-} \equiv { (\vec \sigma_ 3 \cdot \vec q~' )(i \vec \sigma_1 \cdot \vec \sigma_2 \times \vec q~') \over { q'^2 + m_\pi^2} }
{q^2 \over  { q^2 + m_\rho ^2 }} ,
\end{eqnarray}
and
\begin{eqnarray}
F^{II}_{\Delta ^-} \equiv { -(\vec \sigma_ 3  \cdot \vec q~' ) \over { q'^2 + m_\pi^2} }
{{(\vec \sigma_2 \cdot \vec q) \cdot (i \vec \sigma_1 \cdot \vec q \times \vec q~') },
\over  { q^2 + m_\rho ^2 }} ~.
\end{eqnarray}
The  $R^{\pi\rho}_{KR}$, $R^{\pi\rho}_{\Delta^+}$, and  $R^{\pi\rho}_{\Delta^-}$
 form factors are given  in terms of regularization form factors 
 at the meson-baryon-baryon vertices $F_i$
 as \cite{Stadler.1995} 
\begin{eqnarray}
&&R_{KR}^{\pi\rho}(q^2, {q'}^2) = 
 { g^2_\rho g^2 \over 16 m^3} [F_{\rho N N_D}(q^2) 
+\kappa _\rho F_{\rho N N_P}(q^2) ]F_{\rho N N_D}(q^2) F_{\pi NN}^2({q'}^2)
\end{eqnarray}
and
\begin{eqnarray}
&&R_{\Delta^-}^{\pi\rho}(q^2, {q'}^2) = {1 \over 4 } R_{\Delta^+}^{\pi\rho}(q^2, {q'}^2)
\label{formfactor1}
\cr &&
={1 \over 4 } { g _\rho \over 48 m^5 } G^*_{M\rho}{m \over M}
{ 5 M -m \over M-m} m g^* g \cr &&
\times [F_{\rho NN_D}(q^2)+\kappa_\rho F_{\rho NN_P}(q^2)]F_{\rho N\Delta}(q^2)
F_{\pi N\Delta}(q'^2) F_{\pi NN}(q'^2).
\end{eqnarray}
The $F_i$  are taken in monopole form
\begin{equation}
F_i(q^2)=
{\Lambda_i ^2 - m_b^2 \over \Lambda_i^2 + q^2},
\end{equation}
with $i=\{ {\pi NN},{\pi N \Delta},{\rho N N_D},{\rho N N_P}, {\rho N
  \Delta } \}$. 
 The mass of the boson
at the corresponding vertex, $m_b$, is either $m_\pi$ or $m_\rho$ with
exception  of the case $i=\rho N\Delta $ when $m_b=0$.

We would like to have  matrix elements of the three-body force in a
partial wave basis $| p q \alpha \rangle _1$, where 
$p $ and $ q$ are magnitudes of Jacobi momenta 
($\vec  p $ is the relative momentum between particles 2 and 3 and 
$\vec q$ is the momentum of spectator particle 1 relative to the 2-3
pair) 
and $\alpha $ denotes discrete quantum 
numbers which we separate in spin, $\alpha _J$, 
and isospin, $\alpha _T$, parts
\begin{eqnarray}
| p q \alpha \rangle_1  &\equiv& | p (ls) j q (\lambda {1 \over 2} ) I (j,I) JM; (t {1 \over 2} ) T M_T \rangle_1 
\cr &=&
 | p (ls) j q (\lambda {1 \over 2} ) I (j,I) JM \rangle_1 |\alpha_T \rangle_1
\cr &=&
 | p q \alpha_J \rangle_1 |\alpha _T \rangle_1 
\end{eqnarray}
The partial wave states corresponding to different spectator nucleon $i$
($i=1,2,3$) can be obtained from $| p q \alpha \rangle_1$ acting with
 proper permutation operator, 
for instance  $P_{13} P_{23} |pq\alpha\rangle _1 = |p'q'\alpha'\rangle _3 $.

According to the scheme presented in  Ref. \cite{Huber.1994} these 
 matrix elements can be  calculated as 
\begin{eqnarray}
 {_1}\langle p q  \alpha | I_\pm {\cal  F R } (1+P) | p' q' \alpha ' \rangle_ 1 &&=
\sum_{\alpha''} \int_0 ^\infty \int_0^\infty { {p''}^2 dp'' {q''}^2dq''}
{_1}\langle p q \alpha | I_\pm {\cal   F R} |p''q''\alpha''\rangle_3 \cr 
 && \times
{_3}\langle p'' q'' \alpha'' |(1+P)|p'q'\alpha'\rangle_1
\end{eqnarray} 
with
\begin{eqnarray}
{_1}\langle p q \alpha  | I_\pm {\cal  F R} |p''q''\alpha''\rangle_3
&&=
\sum_{ \alpha''' } \int_0 ^\infty \int_0^\infty { {p'''}^2 dp''' {q'''}^2dq'''}
\sum_{\dot {\alpha}} \int_0 ^\infty \int_0^\infty { \dot{p}^2 d\dot{p} \dot{q}^2d \dot{q}} \cr && \times
\sum_{\ddot{\alpha}} \int_0 ^\infty \int_0^\infty { \ddot{p}^2 d\ddot{p} \ddot{q}^2d\ddot {q}}
~~{_1}\langle p q  \alpha  |p'''q'''\alpha'''\rangle_2 \cr
&& \times 
\Bigl(
{_2}\langle p''' q''' \alpha_J'''|{\cal F}^{(2)} {\cal R}^{(2)}|\dot{p} \dot{q} \dot{\alpha}_{J} \rangle_2
~~{_2}\langle \dot{p} \dot{q} \dot{\alpha}_{J}| \ddot{p} \ddot{q} \ddot{\alpha}_{J} \rangle_3 
\cr &&  \times
{_3}\langle \ddot{p} \ddot{q} \ddot{\alpha}_{J} | {\cal F}^{(3)} {\cal R}^{(3)} |p'q'\alpha_J'\rangle_3 
\Bigr) ~~{_2}\langle \alpha'''_T | I_\pm | \alpha_T' \rangle_3,
\end{eqnarray}
where spin operators ${\cal F} $ and form factors $\cal R$ are taken
 among  $F_{KR}$, $F_{\Delta^+}^I $, $F_{\Delta^+}^{II}$,
 $F_{\Delta^-}^I $, 
  $F_{\Delta^-}^{II}$, and 
 $R^{\pi\rho}_{KR}$, $R^{\pi\rho}_{\Delta^+}$ and
 $R^{\pi\rho}_{\Delta^-}$, 
respectively, for different contributing terms.

The matrix elements of the isospin parts appearing in Eq. (\ref{Im}) 
are given \cite{Huber.1994} by 
\begin{eqnarray}
&&{_2}\langle \alpha_T |I_-|\alpha'_T \rangle_3 
=  {_2\langle} (t{1\over 2}) TM_T | \vec \tau_2\cdot \vec \tau_3 | 
(t'{1\over 2}) T'M_{T'} \rangle _3 
\cr 
&& = 
 \delta_{TT'} \delta _{M_T M_{T'}}
(-6)(-)^t\sqrt{\hat t\hat {t'}}
\Biggr\{
\begin{array} {ccc}
 {1 \over 2}  & {1\over 2} &    t' \\ 
 {1 \over 2}  &    1 & {1\over 2} \\
 t & {1\over 2} & T         \\
\end{array}
\Biggl \},
\end{eqnarray}
\begin{eqnarray}
&&{_2}\langle\alpha_T |I_+|\alpha' _T\rangle_3 = 
 {_2\langle } (t{1\over 2}) TM_T | i \vec \tau_1\cdot \vec \tau_2 \times \vec \tau_3 | 
(t' {1\over 2} ) T'M_{T'} \rangle _3
\cr && 
= -
 \delta_{TT'} \delta _{M_T M_{T'}}
24   (-)^{2T} \sqrt{ \hat t \hat {t'} } \sum_{\lambda=1/2} ^{t+1/2} (-)^{3\lambda +1/2} 
\Biggr\{ 
\begin{array}{ccc}
 \lambda & {1 \over 2} & 1 \cr 
                  {1\over 2} & { 1 \over 2} & t 
\end{array} 
\Biggl \} 
\Biggr \{ 
\begin{array}{ccc}
 T & {1\over 2}& t \cr
                   {1\over 2} & 1 & \lambda \cr 
                  t' & {1\over 2} & {1\over 2} 
\end{array} 
\Biggl \}    
\end{eqnarray}
where we use abbreviation $\hat a \equiv 2a+1$.

In the following subsections we will present the resulting expressions
for the partial wave decomposed matrix elements 
 ${_2}\langle p''' q''' \alpha_J'''|{\cal F}^{(2)} {\cal R}^{(2)}|\dot{p} \dot{q} \dot{\alpha}_{J} \rangle_2$ and $
{_3}\langle \ddot{p} \ddot{q} \ddot{\alpha}_{J}  | {\cal F}^{(3)} {\cal R}^{(3)}
|p'q'\alpha_J'\rangle_3 $ 
 of different contributing terms to $\pi - \rho$ Tucson-Melbourne 3NF.

\subsection{The Kroll-Ruderman term $F_{KR} R^{\pi\rho}_{KR} $}
\label{krol-rud}

The matrix elements of 
 ${_2}\langle p''' q''' \alpha_J'''|{\cal F}^{(2)} {\cal R}^{(2)}|\dot p \dot q \dot\alpha_J \rangle_2$ and $
{_3}\langle \ddot{p} \ddot{q} \ddot{\alpha}_{J}  | {\cal F}^{(3)} {\cal R}^{(3)}
|p'q'\alpha_J'\rangle_3 $  for the 
$F_{KR} R_{KR}^{\pi\rho}$ term  are identified as
\begin{eqnarray}
&& {_2}\langle p''' q''' \alpha_J'''|{\cal F}^{(2)} {\cal R}^{(2)}|\dot p \dot q \dot \alpha_{J} \rangle_2 
\to { g^2_\rho g^2 \over 16 m^3}{_2}\langle p''' q''' \alpha_J''' | { -(\vec \sigma_ 3 \cdot \vec q~' ) \over { q'^2 + m_\pi^2} } F_{\pi NN}^2({q'}^2)|\dot p \dot q \dot \alpha_{J} \rangle_2,
\cr &&
 {_3}\langle \ddot p \ddot q \ddot \alpha_J | {\cal F}^{(3)} {\cal R}^{(3)} |p'q'\alpha_J'\rangle_3
\to\cr 
&& {_3}\langle \ddot{p} \ddot{q} \ddot{\alpha}_{J} | { i\vec \sigma_1 \cdot \vec \sigma_2 \times \vec q 
\over { q^2 + m_\rho ^2 }} 
[F_{\rho N N_D}(q^2) +\kappa _\rho F_{\rho N N_P}(q^2) ]F_{\rho N N_D}(q^2)|p'q'\alpha_J'\rangle_3 
\end{eqnarray}
where $\vec q~' = \vec {\dot p} -{\vec p}~'''$ and $\vec q= \vec p~' -\vec {\ddot p}
$. They are given by
\begin{eqnarray}
&&{ g^2_\rho g^2 \over 16 m^3}{_2}\langle p''' q''' \alpha_J''' | { -(\vec \sigma_ 3 \cdot \vec q~' ) 
\over { q'^2 + m_\pi^2} } F_{\pi NN}^2({q'}^2)|\dot p \dot q \dot \alpha_{J} \rangle_2
\cr &&
={ g^2_\rho g^2 \over 16 m^3}
{ \delta ( q'''-\dot q) \over q'''^2} \delta _{j'''\dot j}\delta _{\lambda''' \dot \lambda}
\delta _{I'''\dot I} \delta _{J''' \dot J} \delta _{M'''\dot M} \delta _{\vert l''' -\dot l \vert, 1}
\cr && \times
2 \pi\sqrt{6} (-)^{j'''+1+s'''+\dot s} \sqrt{\hat {s'''} \hat {\dot s}} 
\left\{
\begin{array}{ccc}
{ 1 \over 2} & {1\over2} & \dot s \cr
1 & s''' & {1\over 2} 
\end{array}
\right\}
\left\{
\begin{array}{ccc}
\dot l & \dot s & j''' \cr 
s''' & l''' & 1 
\end{array}
 \right\}
\cr &&
\times
\sqrt{max(l''',\dot l)}(p'''H_{\dot l}^{\pi NN} (p''',\dot p) - \dot p H_{l}^{\pi NN}
(p''',\dot p))(-)^{max(l''',\dot l)}
\end{eqnarray}
and
\begin{eqnarray}
&&{_3}\langle \ddot{p} \ddot{q} \ddot{\alpha}_{J} | { i\vec \sigma_1 \cdot \vec \sigma_2 \times \vec q 
\over { q^2 + m_\rho ^2 }} 
[F_{\rho N N_D}(q^2) +\kappa _\rho F_{\rho N N_P}(q^2) ]F_{\rho N N_D}(q^2)|p'q'\alpha_J'\rangle_3 
\cr && 
= {\delta (\ddot q -q') \over q'^2}
\delta _{\ddot I I'} \delta_{\ddot \lambda  \lambda '} \delta _{\ddot j j'}
\cr &&
\times (-)^{1+\ddot j+s'} 12\sqrt{6}\pi \sqrt{\hat {l'} \hat {\ddot l}} 
\Biggr\{ 
\begin{array}{ccc}
l' & s' & \ddot j \cr \ddot s & \ddot l & 1
\end{array}
 \Biggl\} 
\cr &&
\times \Biggl( 
\begin{array}{ccc}
l' & 1 & \ddot l \cr 
0 & 0 & 0 
\end{array}
 \Biggr)
\sqrt {\hat {s'} \hat {\ddot s} }   
\Biggl\{ 
\begin{array}{ccc}
1 & 1 & 1 \cr {1\over 2} & {1\over 2} & s' \cr 
{1 \over 2} & {1\over 2} & \ddot s 
\end{array} 
\Biggr\} 
\cr && 
\times \Biggl( \ddot p H_{l'}^{KR,\rho NN} ( p', \ddot p) - 
p' H_{\ddot l} ^{KR, \rho NN} ( p', \ddot p) \Biggr) ,
\label{KReq}
\end{eqnarray}
with 
\begin{eqnarray}
&&H_l^{\pi NN}(p''',\dot p)=
{1 \over p''' \dot p} \Bigl( Q_l(B_{m_\pi})-Q_l(B_{\Lambda_{\pi NN}}) \Bigr)
+{\Lambda_{\pi NN} ^2 -m_\pi^2 \over 2(p''' \dot p)^2} Q'_l(B_{\Lambda_{\pi NN}}), 
\cr &&
H_l^{KR,\rho NN}(p',\ddot p) =
{1 \over p' \ddot p } 
\left( Q_{ l}(B_{m_\rho}) - Q_{\bar l}
( B_{\Lambda_{\rho NN_D}}) \right) 
+{ \Lambda_{\rho NN_D}^2 -m_\rho^2 \over 2 (p'\ddot p) ^2 } Q'_{ l} 
(B_{\Lambda_{\rho NN_D}}) 
\cr
&+&
{ \kappa_\rho  \over p' \ddot p } \left( 
    Q_{ l}(B_{m_\rho})
- { \Lambda^2 _{\rho NN_P} - m^2_\rho \over 
    \Lambda^2 _{\rho NN_P} -\Lambda^2_{\rho NN_D}   }
    Q_{ l}(B_{\Lambda_{\rho NN_D}}) 
+  { \Lambda^2 _{\rho NN_D} - m^2_\rho \over 
    \Lambda^2 _{\rho NN_P } -\Lambda ^2_{\rho NN_D} }
    Q_{ l}(B_{\Lambda_{\rho NN_P}})    
 \right)
\cr &&
\label{Hfunc3}
\end{eqnarray}
and
\begin{eqnarray}
&&
B_{m_\pi}= {{p'''}^2+\dot p^2+m_\pi^2 \over 2 p''' \dot p},
~~~ B_{\Lambda_{\pi NN} }= {{p'''}^2+\dot p^2+\Lambda_{\pi NN} ^2 \over 2 p''' \dot p},
\cr && 
B_{m_\rho}= {p'^2+\ddot p^2+m_\rho^2 \over 2 p' \ddot p},~~~
B_{\Lambda_{\rho NN_D}} = {p'^2+\ddot p^2+\Lambda_{\rho N N_D} ^2 \over 2 p' \ddot p},
\cr && 
B_{\Lambda_{\rho NN_P}} = {p'^2+\ddot p^2+\Lambda_{\rho N N_P} ^2 \over 2 p' \ddot p}.
\end{eqnarray}
The $Q_l(x) $ are Legendre functions of the second kind.

\subsection{The term  $F_{\Delta+}^I R_{\Delta+}^{\pi\rho}$}

The term  $F_{\Delta+}^I $ is written as
\begin{eqnarray}
F_{\Delta+}^I = { (\vec \sigma_ 3 \cdot \vec q~' )(\vec \sigma_2 \cdot \vec q~') \over { q'^2 + m_\pi^2} }
{q^2   \over  { q^2 + m_\rho ^2 }} 
=\sum _\mu (-)^\mu \Biggr\{ \sqrt{4\pi \over 3} 
 { q'Y_1 ^{-\mu} (\hat q') (\sigma_ 3 \cdot q' ) \over { q'^2 + m_\pi^2} } \Biggr\}
\Biggr\{ { \sigma_2 ^{\mu} 
 {q}^2  
 \over  { {q}^2 + m_\rho ^2 }}  \Biggl\}
\label{FIDP}
\end{eqnarray}
and the matrix elements of $-\mu$ component of 
 ${_2}\langle p''' q''' \alpha_J'''|{\cal F}^{(2)} {\cal R}^{(2)}|\dot p \dot q \dot\alpha_J \rangle_2$ and  $\mu$ component of
$ {_3}\langle \ddot p \ddot q \ddot \alpha_J | {\cal F}^{(3)} {\cal R}^{(3)} |p'q'\alpha_J'\rangle_3 $ for 
$F_{\Delta+}^I R_{\Delta+}^{\pi\rho}$ are identified as
\begin{eqnarray}
&&\Bigl\{ {_2}\langle p''' q''' \alpha_J''' |{\cal F}^{(2)} {\cal R}^{(2)}|\dot p \dot q \dot\alpha_J \rangle_2 \Bigr\}^{-\mu}
\to \cr && { g _\rho g \over 48 m^5 } G^*_{M\rho}{m \over M}
{ 5 M -m \over M-m} m g^* 
{_2}\langle p''' q''' \alpha_J'''|\sqrt{4\pi \over 3} 
 { q'Y_1 ^{-\mu} (\hat q') (\sigma_ 3 \cdot q' ) \over { q'^2 + m_\pi^2} } 
F_{\pi NN} (q'^2)F_{\pi N\Delta}(q'^2)
|\dot p \dot q \dot\alpha_J \rangle_2 ,
\cr &&
\Bigl\{ {_3}\langle \ddot p \ddot q \ddot \alpha_J | {\cal F}^{(3)} {\cal R}^{(3)} |p'q'\alpha_J'\rangle_3 \Bigr\}^\mu 
\to\cr 
&&{_3}\langle \ddot p \ddot q \ddot \alpha_J |
 { \sigma_2 ^{\mu} 
 {q}^2  
 \over  { {q}^2 + m_\rho ^2 }} [F_{\rho NN_D}(q^2)+\kappa_\rho F_{\rho NN_P}(q^2)]F_{\rho N\Delta}(q^2) |p'q'\alpha_J'\rangle_3,
\label{eDpI}
\end{eqnarray}
where $\vec q~' = \vec {\dot p} -{\vec p}~'''$ and $\vec q= \vec p~' -\vec {\ddot p}$. They are given by
\begin{eqnarray}
&&
{_2}\langle p''' q''' \alpha_J''' \vert \sqrt{4\pi \over 3}
 {      q'Y_1 ^{-\mu} (\hat q') (\sigma_ 3 \cdot \vec q~' ) \over q'^2 + m_\pi^2}  
F_{\pi NN} (q'^2) F_{\pi N\Delta} (q'^2)  
\vert \dot p \dot q \dot\alpha_J \rangle _2
\cr && 
= 
{\delta (q'''- \dot q) \over {q'''}^2 } \delta _{\lambda ''' \dot \lambda } \delta _{ I''' \dot I}
C(1 -\mu \dot J \dot M , J''' M''' )\cr
&&
\times (-)^{I'''+J'''} \sqrt{ \hat j''' \hat {\dot j} \hat s''' \hat {\dot s} \hat {\dot J}}
\left\{
\begin{array}{ccc}
{1\over 2} & {1\over2} & \dot s \cr 
1 & s''' &  {1\over 2} \cr 
\end{array}
 \right\}
\left\{
\begin{array}{ccc}
1 & \dot j & j''' \cr I''' & J''' & \dot J \cr 
\end{array}
 \right\}
\cr
&&
\times \Biggl[ \delta _{l''' \dot l} {2\pi \over 3 } \sqrt{6} (-)^{l''' +1} 
\tilde H^{\pi NN-\pi N\Delta}_{l'''} (p''',\dot p) 
\left\{
\begin{array}{ccc}
j''' & \dot j & 1 \cr \dot s & s''' & l''' \cr 
\end{array}
\right\}
\cr
&&
-40 \pi \sqrt{6} (-)^{s'''+\dot j} 
\left\{
\begin{array}{ccc}
2 & 1 & 1\cr \dot l & \dot s & \dot j \cr l'''& s'''& j'''\cr 
\end{array}
\right\}
\cr 
&&
\times \sum _{\bar l} \hat {\bar l} 
H_{\bar l}^{\pi NN-\pi N\Delta} (p''',\dot p) \sum_{a+b=2} 
{ {p'''}^a {\dot p}^b \over \sqrt{(2a)!(2b)!} } 
\left\{
\begin{array}{ccc}
 b & a & 2 \cr l''' & \dot l & \bar l \cr 
\end{array}
 \right\}
C(a0\bar l0,l'''0 )C(b 0 \bar l 0 , \dot l 0) 
\Biggr]
\label{pRI}
\end{eqnarray}
and
\begin{eqnarray}
&&
_3\langle \ddot p \ddot q \ddot \alpha_J \vert \sigma_2 ^{\mu} {q^2 \over q^2 + m_\rho^2 }
[F_{\rho NN_D}(q^2)+\kappa_\rho F_{\rho NN_P}(q^2)]F_{\rho N\Delta}(q^2)
 \vert p' q '\alpha' \rangle_3
\cr &&
= { \delta (\ddot q-q') \over q'^2 } \delta _{\ddot l l'} \delta _{ \ddot \lambda \lambda'}
\delta_{\ddot I I'} \sqrt{6}~2\pi 
\tilde H_{ l'} ^{ \rho NN} (\ddot p, p') \cr
&&
\times (-)^{1+l'+s'+\ddot s+I'+\ddot J}
\sqrt{\hat j' \hat {\ddot j} \hat s' \hat {\ddot s} \hat J'}
\left\{
\begin{array}{ccc}
s' & j' & l' \cr \ddot j & \ddot s & 1\cr
\end{array}
\right\}
\left\{
\begin{array}{ccc}
{1\over2}&{1\over2}& s' \cr 1 & \ddot s  & {1\over 2}\cr
\end{array}
\right\}
\cr &&
\times 
\left\{
\begin{array}{ccc}
1 &j' & \ddot j \cr I' & \ddot J & J' \cr
\end{array}
 \right\}
C(1 \mu J' M', \ddot J \ddot M),
\label{FI-J3}
\end{eqnarray}
with
\begin{eqnarray}
&&\tilde H_{l} ^{\pi NN-\pi N\Delta}(p''',\dot p) 
= { (\Lambda _{\pi NN}^2-m_\pi^2 )(\Lambda_{\pi N\Delta}^2-m_\pi^2)\over p''' \dot p }
\cr && \times
 \Biggr(
{m_{\pi}^2 \over (\Lambda_{\pi N\Delta}^2 - m_\pi^2)(m_\pi^2 -\Lambda_{\pi NN}^2)} 
Q_{ l} (B_{m_{\pi}}) 
\cr &&
+
{\Lambda_{\pi NN}^2 \over (m_\pi^2 -\Lambda_{\pi NN} ^2)(\Lambda_{\pi NN}^2 -\Lambda_{\pi N\Delta}^2)}
 Q_{ l} (B_{\Lambda_{\pi NN}})
\cr && 
+{\Lambda_{\pi N\Delta}^2 \over (\Lambda_{\pi NN}^2 - \Lambda_{\pi N\Delta} ^2)(\Lambda_{\pi N\Delta}^2 -m_\pi^2)}
Q_{ l} (B_{\Lambda_{\pi N\Delta}})  
\Biggr) ,
\end{eqnarray}
\begin{eqnarray}
&& H_{\bar l}^{\pi NN-\pi N\Delta}(p''',\dot p) 
\cr
&&= {1\over p''' \dot p } \Biggr(
Q_{\bar l}(B_{m_\pi})
+{\Lambda_{\pi N\Delta}^2-m_\pi^2 \over \Lambda_{\pi NN}^2 -\Lambda_{\pi N\Delta}^2}
Q_{\bar l}(B_{\Lambda_{\pi NN}}) +
{\Lambda_{\pi NN}^2-m_\pi^2 \over \Lambda_{\pi N\Delta}^2 -\Lambda_{\pi NN}^2}
Q_{\bar l} (B_{\Lambda_{\pi N\Delta}})
\Biggr) 
\end{eqnarray}
and
\begin{eqnarray}
&&\tilde H_{\bar l} ^{\rho NN}(p', \ddot p) 
\cr 
&& 
= { ( \Lambda_{\rho NN_D}^2 -m_\rho^2 ) \Lambda_{\rho N\Delta} ^2  \over p' \ddot p }
\cr && 
\times 
\Biggr( 
{ m_\rho ^2 \over ( \Lambda_{\rho N \Delta} ^2 - m_{\rho}^2 )  
(m_\rho^2 -\Lambda_{\rho NN_D} ^2) }  Q_{\bar l} ( B_{m_\rho}) 
+ { \Lambda_{\rho NN_D} ^2  \over ( m_\rho^2  - \Lambda_{\rho NN_D}^2 ) 
(\Lambda_{\rho NN_D}^2 - \Lambda_{\rho N\Delta} ^2) } Q_{\bar l} (B_{\Lambda_{\rho NN_D}}) 
\cr && 
+ { \Lambda_{\rho N\Delta} ^2  \over ( \Lambda_{\rho NN_D} ^2  - \Lambda_{\rho N\Delta}^2 ) 
(\Lambda_{\rho N\Delta} ^2- m_\rho^2) } Q_{\bar l} (B_{\Lambda_{\rho N\Delta}}) \Biggr)
\cr &&
+ {(\kappa_\rho) ( \Lambda_{\rho NN_P}^2 -m_\rho^2 ) \Lambda_{\rho N\Delta} ^2  \over p' \ddot p }
\cr && 
\times \Biggr( 
{ m_\rho ^2 \over ( \Lambda_{\rho N \Delta} ^2 - m_{\rho}^2 )  
(m_\rho^2 -\Lambda_{\rho NN_P} ^2) }  Q_{\bar l} ( B_{m_\rho}) 
+ { \Lambda_{\rho NN_P} ^2  \over ( m_\rho^2  - \Lambda_{\rho NN_P}^2 ) 
(\Lambda_{\rho NN_P}^2 - \Lambda_{\rho N\Delta} ^2) } Q_{\bar l} (B_{\Lambda_{\rho NN_P}}) 
\cr && 
+ { \Lambda_{\rho N\Delta} ^2  \over ( \Lambda_{\rho NN_P} ^2  - \Lambda_{\rho N\Delta}^2 ) 
(\Lambda_{\rho N\Delta}^2 - m_\rho^2) } Q_{\bar l} (B_{\Lambda_{\rho N\Delta}}) \Biggr).
\cr &&
\label{Hfunc2}
\end{eqnarray}
The $B_{\Lambda_{\pi N\Delta} }$ and $B_{\Lambda_\rho N \Delta}$ are given by
\begin{eqnarray}
 && B_{\Lambda_{\pi N\Delta} }= {{p'''}^ 2+\dot p^2+\Lambda_{\pi N\Delta} ^2 \over 2 p''' \dot p}, ~~~
 B_{\Lambda_{\rho N\Delta} }= {{p'''}^ 2+\dot p^2+\Lambda_{\rho N\Delta} ^2 \over 2 p''' \dot p}.
\end{eqnarray}

The summation over $\mu$ in Eq. (\ref{FIDP}) can be carried through
resulting in
\begin{equation}
\sum_{\mu} (-)^\mu C(1 ~-\mu \dot J \dot M,J M)C(1 ~\mu J'M',\dot J\dot M)=\delta_{JJ'}\delta_{MM'}(-)^{\dot J-J}\sqrt{\hat {\dot J} \over \hat J}.
\end{equation}

\subsection{The term  $F_{\Delta +} ^{II}R_{\Delta+}^{\pi\rho}$ }

The term  $F_{\Delta+}^{II} $ is written  as
\begin{eqnarray}
&& F_{\Delta+}^{II} = { (\vec \sigma_ 3 \cdot \vec q~' ) \over { q'^2 + m_\pi^2} }(\vec q\cdot \vec q~')
{(\vec \sigma_2\cdot \vec q)
 \over  { q^2 + m_\rho ^2 }}\cr && 
=\sum _\mu (-)^\mu \Biggr\{ \sqrt{4\pi \over 3} 
 { q'Y_1 ^{-\mu} (\hat q') (\vec \sigma_ 3 \cdot \vec q~' ) \over { q'^2 + m_\pi^2} } \Biggr\}
\Biggr\{ \sqrt{4\pi \over 3} 
 { qY_1 ^{\mu} (\hat q) (\vec \sigma_ 2 \cdot \vec q ) \over { q^2 + m_\rho ^2} } \Biggr\}.
\label{FIIDP}
\end{eqnarray}
The matrix elements of $-\mu$ component of 
 ${_2}\langle p''' q''' \alpha_J'''|{\cal F}^{(2)} {\cal R}^{(2)}|\dot p \dot q \dot\alpha_J \rangle_2$ and  $\mu$ component of
$ {_3}\langle \ddot p \ddot q \ddot \alpha_J | {\cal F}^{(3)} {\cal R}^{(3)} |p'q'\alpha_J'\rangle_3 $
for 
$F_{\Delta+}^{II} R_{\Delta+}^{\pi\rho}$ are identified as
\begin{eqnarray}
&&\Bigl\{ {_2}\langle p''' q''' \alpha_J''' |{\cal F}^{(2)} {\cal R}^{(2)}|\dot p \dot q \dot\alpha_J \rangle_2 \Bigr\}^{-\mu}
\to \cr && { g _\rho g \over 48 m^5 } G^*_{M\rho}{m \over M}
{ 5 M -m \over M-m} m g^* 
{_2}\langle p''' q''' \alpha_J''' |\sqrt{4\pi \over 3} 
 { q'Y_1 ^{-\mu} (\hat q') (\vec \sigma_ 3 \cdot \vec q~' ) \over { q'^2 + m_\pi^2} } 
F_{\pi NN} (q'^2)F_{\pi N\Delta}(q'^2)
|\dot p \dot q \dot\alpha_J \rangle_2 ,
\cr &&
\Bigl\{ {_3}\langle \ddot p \ddot q \ddot \alpha_J | {\cal F}^{(3)} {\cal R}^{(3)} |p'q'\alpha_J'\rangle_3 \Bigr\}^\mu 
\to\cr 
&&{_3}\langle \ddot p \ddot q \ddot \alpha_J |\sqrt{4\pi \over 3} 
 { q Y_1 ^{\mu} (\hat q) (\vec \sigma_ 3 \cdot \vec q ) \over { q^2 + m_\rho ^2} } 
 [F_{\rho NN_D}(q^2)+\kappa_\rho F_{\rho NN_P}(q^2)]F_{\rho N\Delta}(q^2) |p'q'\alpha_J'\rangle_3,
\label{eRII}
\end{eqnarray}
where $\vec q~' = \vec {\dot p} -{\vec p}~'''$ and $\vec q= \vec p~' -\vec {\ddot p}$. 
The first term in Eq.(\ref{eRII}) is equal to  Eq. (\ref{pRI}) and the 
second  is given by
\begin{eqnarray}
&&{_3}\langle\ddot p \ddot q \ddot \alpha_J|\sqrt{4\pi \over 3} 
 { q Y_1 ^{\mu} (\hat q) (\vec \sigma_ 3 \cdot \vec q ) \over { q^2 + m_\rho ^2} } 
 [F_{\rho NN_D}(q^2)+\kappa_\rho F_{\rho NN_P}(q^2)]F_{\rho N\Delta}(q^2) |p'q'\alpha_J'\rangle_3
\cr && 
= 
{\delta (q'- \ddot q) \over q'^2 } \delta _{\lambda'  \ddot \lambda } \delta _{ I' \ddot I}
C(1 ~\mu J' M' , \ddot J  \ddot M )\cr
&&
\times (-)^{I'+\ddot J+s'-\ddot s} \sqrt{ \hat j' \hat {\ddot j} \hat s' \hat {\ddot s} \hat J'}
\left\{
\begin{array}{ccc}
{1\over 2} & {1\over2} & \ddot s \cr 
1 & s' &  {1\over 2} \cr 
\end{array}
 \right\}
\left\{
\begin{array}{ccc}
1 & \ddot j & j' \cr I' & J' & \ddot J \cr 
\end{array}
 \right\}
\cr
&&
\times \Biggl[ \delta _{l' \ddot l} {2\pi \over 3 } \sqrt{6} (-)^{l' +1} 
\tilde H^{\rho NN}_{l'} (p',\ddot p) 
\left\{
\begin{array}{ccc}
j' & \ddot j & 1 \cr \ddot s & s' & l' \cr 
\end{array}
\right\}
\cr
&&
-40 \pi \sqrt{6} (-)^{s'+\ddot j} 
\left\{
\begin{array}{ccc}
2 & 1 & 1\cr \ddot l & \ddot s & \ddot j \cr l'& s'& j'\cr 
\end{array}
\right\}
\cr 
&&
\times \sum _{\bar l} \hat {\bar l} 
H_{\bar l}^{\rho NN} (p',\ddot p) \sum_{a+b=2} 
{ {p'}^a {\ddot p}^b \over \sqrt{(2a)!(2b)!} } 
\left\{
\begin{array}{ccc}
 b & a & 2 \cr l' & \ddot l & \bar l \cr 
\end{array}
 \right\}
C(a0\bar l0,l'0 )C(b 0 \bar l 0 , \ddot l 0) 
\Biggr]
\label{pRII}
\end{eqnarray}
with 
\begin{eqnarray}
&&H ^{\rho NN} _{\bar l} (p' ,\ddot p) 
\cr
&&
={ 1   \over p' \ddot p } \Biggl( 
{\Lambda_{\rho N\Delta} ^2 \over \Lambda_{\rho N\Delta}^2 - m_\rho^2 }
    Q_{\bar l}(B_{m_\rho})
+ { \Lambda^2 _{\rho N\Delta}  \over \Lambda^2_{\rho NN_D}  - \Lambda^2 _{\rho N\Delta}  }
    Q_{\bar l}(B_{\Lambda_{\rho NN_D}}) 
\cr && 
+  { \Lambda^2 _{\rho N N_D} - m^2_\rho \over 
    \Lambda^2 _{\rho N\Delta} - m_\rho^2 }
{\Lambda_{\rho N\Delta} ^2 \over 
    \Lambda^2 _{\rho N\Delta} -\Lambda^2_{\rho NN_D}   }
    Q_{\bar l}(B_{\Lambda_{\rho N\Delta}})    
 \Biggl)
\cr
&+&
{ \kappa_\rho  \over p' \ddot p } \Biggl( 
{\Lambda_{\rho N\Delta} ^2 \over \Lambda_{\rho N\Delta}^2 - m_\rho^2 }
    Q_{\bar l}(B_{m_\rho})
+ { \Lambda^2 _{\rho N\Delta}  \over \Lambda^2_{\rho NN_P}  - \Lambda^2 _{\rho N\Delta}  }
    Q_{\bar l}(B_{\Lambda_{\rho NN_P}}) 
\cr && 
+  { \Lambda^2 _{\rho N N_P} - m^2_\rho \over 
    \Lambda^2 _{\rho N\Delta} - m_\rho^2 }
{\Lambda_{\rho N\Delta} ^2 \over 
    \Lambda^2 _{\rho N\Delta} -\Lambda^2_{\rho NN_P}   }
    Q_{\bar l}(B_{\Lambda_{\rho N\Delta}})    
 \Biggl).
\cr &&
\label{Hfunc}
 \end{eqnarray}

\subsection{The term  $F_{\Delta -}^I R_{\Delta -}^{\pi\rho}$ }

Using the identity 
\begin{eqnarray}
\vec \sigma_1 \times \vec \sigma_2 \cdot \vec q~' = i \sqrt{2} \sqrt{ 4\pi \over 3}
q' \sum_\mu (-)^\mu \sigma_2^\mu \{ \vec \sigma_1, Y_1(\hat q~') \}^{1,-\mu},
\label{sigsigQ}
\end{eqnarray}
the term  $F_{\Delta-}^{I}$ is written  as 
\begin{eqnarray}
&&F^I_{\Delta ^-} \equiv { (\vec \sigma_ 3 \cdot \vec q~' )(i \vec \sigma_1 \cdot \vec \sigma_2 \times \vec q~') 
\over { q'^2 + m_\pi^2} }
{q^2 \over  { q^2 + m_\rho ^2 }} 
\cr &&
= \sqrt{2}\sqrt{4\pi \over 3} \sum_\mu  (-)^\mu
\Bigl\{ { (\vec \sigma_ 3 \cdot q' ) q'\{ \vec  \sigma_1, Y_1 (\hat q~') \}^{1,-\mu} \over { {q'}^2 + m_\pi^2} }
\Bigr\}
\Bigl\{
{\sigma_2 ^\mu {q}^2 \over  { {q}^2 + m_\rho ^2 }} \Bigr\} .
\end{eqnarray}
The matrix elements of $-\mu$ component of 
 ${_2}\langle p''' q''' \alpha_J'''|{\cal F}^{(2)} {\cal R}^{(2)}|\dot p \dot q \dot\alpha_J \rangle_2$ and  $\mu$ component of
$ {_3}\langle \ddot p \ddot q \ddot \alpha_J | {\cal F}^{(3)} {\cal R}^{(3)} |p'q'\alpha_J'\rangle_3 $
for 
$F_{\Delta-}^{I} R_{\Delta-}^{\pi\rho}$ are identified as
\begin{eqnarray}
&&\Bigl\{ {_2}\langle p''' q''' \alpha_J''' |{\cal F}^{(2)} {\cal R}^{(2)}|\dot p \dot q \dot\alpha_J \rangle_2 \Bigr\}^{-\mu}
\to \cr && {1 \over 4} { g _\rho g \over 48 m^5 } G^*_{M\rho}{m \over M}
{ 5 M -m \over M-m} m g^* 
\cr &&
\times
{_2}\langle p''' q''' \alpha_J'''|\sqrt{2}\sqrt{4\pi \over 3} 
 { q' \{\vec \sigma_1, Y_1  (\hat q')\}^{1,-\mu} (\vec \sigma_ 3 \cdot \vec q~' ) \over { q'^2 + m_\pi^2} } 
F_{\pi NN} (q'^2)F_{\pi N\Delta}(q'^2)
|\dot p \dot q \dot\alpha_J \rangle_2 ,
\cr &&
\Bigl\{ {_3}\langle \ddot p \ddot q \ddot \alpha_J | {\cal F}^{(3)} {\cal R}^{(3)} |p'q'\alpha_J'\rangle_3 \Bigr\}^\mu 
\to\cr 
&&{_3}\langle \ddot p \ddot q \ddot \alpha_J |
 { \sigma_2 ^{\mu} 
 {q}^2  
 \over  { {q}^2 + m_\rho ^2 }} [F_{\rho NN_D}(q^2)+\kappa_\rho F_{\rho NN_P}(q^2)]F_{\rho N\Delta}(q^2) |p'q'\alpha_J'\rangle_3,
\label{eRIm}
\end{eqnarray}
where $\vec q~' = \vec {\dot p} -{\vec p}~'''$ and $\vec q= \vec p~' -\vec {\ddot p}$. 
The last term is identical to the last term in  Eq. (\ref{eDpI}) for 
 $F_{\Delta +}^{I}$. The matrix element in the first term of
Eq.(\ref{eRIm}) is given by
\begin{eqnarray}
&&{_2}\langle p''' q''' \alpha_J'''|\sqrt{2}\sqrt{4\pi \over 3} 
 { q' \{\vec \sigma_1, Y_1  (\hat q')\}^{1,-\mu} (\vec \sigma_ 3 \cdot \vec q~' ) \over { q'^2 + m_\pi^2} } 
F_{\pi NN} (q'^2)F_{\pi N\Delta}(q'^2)
|\dot p \dot q \dot\alpha_J \rangle_2 
\cr && 
=(-){ \delta (q'''-\dot q) \over q'''~^2 } \delta _{\lambda''' \dot \lambda} 
\delta _{I''' \dot I} (-)^{I'''+J'''+1} \sqrt{\hat j''' \hat {\dot j} \hat s''' \hat {\dot s} \hat {\dot J} }
C(1-\mu \dot J \dot M, JM) 
\left\{
\begin{array}{ccc}
1&\dot j & j '''\cr I''' & J''' &\dot J \cr
\end{array}
\right\}
\cr
&&
\times \Biggl[  \delta _{l'''\dot l}  4\pi \sqrt{6} (-)^{l'''+s'''+1}
\tilde H_{l'''} ^{\pi NN-\pi N\Delta} (p''',\dot p)
\left\{
\begin{array}{ccc}
l''' & s''' & j''' \cr 1 &\dot j & \dot s \cr 
\end{array}
 \right\}
\left\{
\begin{array}{ccc}
1& 1& 1\cr {1\over 2} & {1 \over 2} & \dot s \cr 
{1\over 2} & {1 \over 2 } & s''' \cr 
\end{array}
 \right\}
 \cr 
&&
+  240 \pi \sqrt{6} (-)^{\dot j} \sum _{\bar l} \hat {\bar l}
 H_{\bar l} ^{\pi NN-\pi N\Delta} (p''',\dot p)
 \sum_{a+b=2}{ p'''~^a~ {\dot p}^b \over \sqrt{(2a)!(2b)!}}
\cr &&
\times
\left\{
\begin{array}{ccc}
a &b & 2 \cr \dot l & l''' & \bar l \cr 
\end{array}
\right\} 
C(a0\bar l 0,l'''0) C(b0\bar l 0,\dot l 0)
\cr &&
\times
\sum_{k}\hat k  
\left\{
\begin{array}{ccc}
2 & k & 1 \cr 1 & 1 & 1 \cr
\end{array}
 \right\}
\left\{
\begin{array}{ccc}
2& k & 1 \cr \dot l & \dot s & \dot j \cr l''' & s''' & j''' \cr
\end{array}
 \right\}
\left\{
\begin{array}{ccc}
1 & 1 & k \cr {1\over2}& {1\over2} & \dot s \cr 
{1\over 2} & {1\over2} & s''' \cr
\end{array}
 \right\}
  \Biggr].
\label{mRI}
\end{eqnarray}

\subsection{The term  $F_{\Delta -}^{II}R_{\Delta-}^{\pi\rho}$ }

Using the identity
\begin{eqnarray}
\vec \sigma_1 \cdot \vec  q \times \vec q~' = -i \sqrt{2} {4\pi \over 3} qq'\sum_\mu (-)^\mu
\{\sigma_1, Y_1(\hat q) \}^{1,-\mu} Y_1 ^\mu (\hat q~'),
\end{eqnarray}  
the term  $F_{\Delta-}^{II}$ is written as 
\begin{eqnarray}
&&F^{II}_{\Delta ^-} \equiv { -(\vec \sigma_ 3 \cdot \vec q~' ) \over { q'^2 + m_\pi^2} }
{{(\vec \sigma_2 \cdot \vec q) (i \vec \sigma_1 \cdot \vec q \times \vec q~') }
\over  { q^2 + m_\rho ^2 }} \cr
&&=- \sqrt{2} {4\pi \over 3} \sum_\mu (-)^\mu
\Bigl\{  { (\vec \sigma_ 3 \cdot \vec q~' ) q' Y_1 ^\mu(\hat q~')
\over { q'^2 + m_\pi^2} }
\Bigr\}
\Bigl\{
{ { (\vec \sigma_2 \cdot \vec q) q \{\vec \sigma_1 ,Y_1^\mu (\hat q) \} ^{1,-\mu} } \over  { q^2 + m_\rho ^2 } }
\Bigr\} .
\end{eqnarray}
The matrix elements of $-\mu$ component of 
 ${_2}\langle p''' q''' \alpha_J'''|{\cal F}^{(2)} {\cal R}^{(2)}|\dot p \dot q \dot\alpha_J \rangle_2$ and  $\mu$ component of
$ {_3}\langle \ddot p \ddot q \ddot \alpha_J | {\cal F}^{(3)} {\cal R}^{(3)} |p'q'\alpha_J'\rangle_3 $
 for 
$F_{\Delta-}^{II} R_{\Delta-}^{\pi\rho}$ are identified as
\begin{eqnarray}
&&\Bigl\{ {_2}\langle p''' q''' \alpha_J''' |{\cal F}^{(2)} {\cal R}^{(2)}|\dot p \dot q \dot\alpha_J \rangle_2 \Bigr\}^{-\mu}
\to \cr &&  -{1 \over 4} { g _\rho g \over 48 m^5 } G^*_{M\rho}{m \over M}
{ 5 M -m \over M-m} m g^* 
{_2}\langle p''' q''' \alpha_J''' |\sqrt{4\pi \over 3} 
 { (\vec \sigma_ 3 \cdot \vec q~' ) q' 
 Y_1 ^{-\mu}  (\hat q')  \over { q'^2 + m_\pi^2} } 
F_{\pi NN} (q'^2)F_{\pi N\Delta}(q'^2)
|\dot p \dot q \dot\alpha_J \rangle_2 ,
\cr &&
\Bigl\{ {_3}\langle \ddot p \ddot q \ddot \alpha_J | {\cal F}^{(3)} {\cal R}^{(3)} |p'q'\alpha_J'\rangle_3 \Bigr\}^\mu 
\to\cr 
&& \sqrt{2} \sqrt{4\pi \over 3 }{_3}\langle \ddot p \ddot q \ddot \alpha_J |
{ { (\vec \sigma_2  \cdot \vec q ) q  \{ \vec \sigma_1, Y_1 (\hat q) \}^{1,\mu} } 
 \over  { {q}^2 + m_\rho ^2 }} [F_{\rho NN_D}(q^2)+\kappa_\rho 
F_{\rho NN_P}(q^2)]F_{\rho N\Delta}(q^2) |p'q'\alpha_J'\rangle_3,\cr
&& 
\label{eRIIm}
\end{eqnarray}
where $\vec q~' = \vec {\dot p} -{\vec p}~'''$ and $\vec q= \vec p~' -\vec {\ddot p}$. 
 They are given by
 \begin{eqnarray}
&&
~{_2}\langle p''' q''' \alpha_J''' \vert \sqrt{ 4\pi \over 3}  {  (\vec \sigma_3 \cdot \vec q~') q'  Y_1^{-\mu}(\hat q~')
 \over {q'}^2 + m_\pi^2 }F_{\pi NN}({q'}^2)F_{\pi N \Delta} ({q'}^2)
 \vert \dot p \dot q \dot\alpha_J \rangle_2
\cr && 
=
{ \delta (q'''-\dot q) \over q'''~^2 } \delta _{\lambda ''' \dot \lambda } \delta _{ I''' \dot I } 
C( 1 ~-\mu \dot J \dot M,J''' M''')  \cr 
&&
\times (-)^{ I'''+J'''} \sqrt{ \hat j''' \hat {\dot j} 
\hat s''' \hat {\dot s} \hat {\dot J} } 
\left\{
\begin{array}{ccc}
{1 \over 2} & {1 \over 2} & \dot s \cr 
                 1 & s''' & {1 \over 2 } \cr 
\end{array}
 \right\}
\left\{
\begin{array}{ccc}
1 & \dot j & j''' \cr I''' & J''' & \dot J \cr
\end{array}
 \right\}
\cr
&& \times
\Biggl[ \delta _{l''' \dot l } {2 \pi \over 3 } \sqrt{6}(-)^{l''' + 1} 
\tilde H _{ l} ^{\pi NN-\pi N\Delta} (p''',\dot p ) 
\left\{
\begin{array}{ccc}
j''' & \dot j & 1 \cr \dot s & s''' & l''' \cr 
\end{array}
 \right\}
\cr 
&& 
- 40 \pi \sqrt{6} (-)^{s'''+\dot j} 
\left\{ 
\begin{array}{ccc}
2 & 1 & 1 \cr 
                \dot l & \dot s & \dot j \cr 
                 l''' & s'''  & j''' \cr 
\end{array}
\right\}
\cr
&& 
\times \sum_{\bar l} \hat {\bar l} H_{\bar l} ^{ \pi NN-\pi N\Delta} (p''',\dot p ) 
\sum_{a+b=2} { {p'''}^a ~{\dot p} ^b \over \sqrt{(2a)! (2b)!} } 
\left\{
\begin{array}{ccc}
 b & a & 2 \cr l''' & \dot l & \bar l \cr 
\end{array}
 \right\}
C(a0\bar l 0, l''' 0) C(b0 \bar l 0 ,\dot l 0) \Biggr] 
\cr &&
\label{mRII}
 \end{eqnarray}
and 
\begin{eqnarray}
&& \sqrt{2} \sqrt{4\pi \over 3 }{_3}\langle \ddot p \ddot q \ddot \alpha_J |
{ { (\vec \sigma_2  \cdot \vec q ) q  \{ \vec \sigma_1, Y_1 (\hat q) \}^{1,\mu} } 
 \over  { {q}^2 + m_\rho ^2 }} [F_{\rho NN_D}(q^2)+\kappa_\rho 
F_{\rho NN_P}(q^2)]F_{\rho N\Delta}(q^2) |p'q'\alpha_J'\rangle_3
\cr && 
=
{ \delta (q'-\ddot q) \over {q'}^2 } \delta _{\lambda' \ddot \lambda} 
\delta _{I' \ddot I} 
\cr && 
\times
(-)^{I'+\ddot J +l'+\ddot l+1} \sqrt{\hat j' \hat {\ddot j} \hat s' \hat {\ddot s} \hat J' }
C(1 ~\mu J'M',\ddot J \ddot M) 
\left\{
\begin{array}{ccc}
1&\ddot j &  j'\cr I' & J' &\ddot J\cr
\end{array}
\right\}
\cr
&&
\times \Biggl[  \delta _{l'\ddot l}  4\pi \sqrt{6} (-)^{l'+s'}
\left\{
\begin{array}{ccc}
l' & s' & j' \cr 1 &\ddot j & \ddot s \cr 
\end{array}
 \right\}
\left\{
\begin{array}{ccc}
1& 1& 1\cr {1\over 2} & {1 \over 2} & \ddot s \cr 
{1\over 2} & {1 \over 2 } & s' \cr 
\end{array}
 \right\}
\tilde H_{l'} ^{\rho NN} (p',\ddot p) \cr 
&&
+  240 \pi \sqrt{6} (-)^{\ddot j} \sum _{k} (-)^{k} \hat k 
\left\{
\begin{array}{ccc}
2 & k & 1 \cr 1 & 1 & 1 \cr
\end{array}
 \right\}
\left\{
\begin{array}{ccc}
2& k & 1 \cr \ddot l & \ddot s & \ddot j \cr l' & s' & j' \cr
\end{array}
 \right\}
\left\{
\begin{array}{ccc}
1 & 1 & k \cr {1\over2}& {1\over2} & \ddot s \cr 
{1\over 2} & {1\over2} & s' \cr
\end{array}
 \right\}
\cr 
&&
\times 
\sum_{a+b=2} { {p'}^a~ {\ddot p}^b\over \sqrt{(2a)!(2b)!}}
\sum_{\bar l}{\hat {\bar l}}
H_{\bar l} ^{\rho NN} (p',\ddot p) 
\left\{
\begin{array}{ccc}
a &b & 2 \cr \ddot l & l' & \bar l \cr 
\end{array}
\right\} 
C(a0\bar l 0,l'0) C(b0\bar l 0,\ddot l 0)   \Biggr].
\end{eqnarray}

Note, that in the case of $\Lambda_{\pi N\Delta} = \Lambda_{\pi NN}$
(as was assumed in \cite{Stadler.1995} and also used 
by us in applications shown in this study), the function  
$H^{\pi NN-\pi N\Delta}_l(p''',\dot p)$ is equal to $H^{\pi NN}_l(p''',\dot p)$
defined in Eq.(\ref{Hfunc3}) and the function 
$\tilde H_l^{\pi NN-\pi N\Delta}(p''',\dot p) $ is equal to  $\tilde
H_l^{\pi NN}(p''',\dot p) $ given by 
\begin{eqnarray}
&&\tilde H_{\bar l} ^{\pi NN}(p''',\dot p) 
\cr
&&= - { m_\pi^2 \over p''' \dot p } \left(
Q_{\bar l} (B_{m_\pi}) - Q_{\bar l} (B_{\Lambda_{\pi NN}}) \right) 
-{ \Lambda^2_{\pi NN} - m_\pi^2 \over 2 (p''' \dot p)^2 } \Lambda^2 _{\pi NN} 
Q'_{\bar l}(B_{\Lambda_{\pi NN}})  
\end{eqnarray}

\end{document}